\documentclass[11pt,showpacs,preprintnumbers,amsmath,amssymb,aps,prd,preprint,nofootinbib]{revtex4}
\pdfoutput=1

\usepackage{amsmath}
\usepackage{amssymb}
\usepackage{latexsym}
\usepackage{amsfonts}
\usepackage{graphicx}
\usepackage{epstopdf}
\usepackage{xcolor}

\usepackage{color}

\usepackage[colorlinks,urlcolor=blue,citecolor=blue]{hyperref}
\def\hhref#1{\href{http://arxiv.org/abs/#1}{arXiv:#1}} 
\def\mc{\mathcal}

\newcommand{\E}{{\mathcal E}}

\newcommand{\be}{\begin{equation}}
\newcommand{\ee}{\end{equation}}
\newcommand{\bear}{\begin{eqnarray}}
\newcommand{\eear}{\end{eqnarray}}
\newcommand{\nn}{\nonumber\\}
\newcommand{\ba}{\begin{array}}
\newcommand{\ea}{\end{array}}

\begin{document}

\title{Hydrodynamics, resurgence and trans-asymptotics}

\author{G\"ok\c ce Ba\c sar}
\email{gbasar@umd.edu}
\affiliation{Maryland Center for Fundamental Physics, University of Maryland, College Park, MD, 20742}

\author{Gerald~V.~Dunne}
\email{gerald.dunne@uconn.edu}
\affiliation{Department of Physics, University of Connecticut, Storrs CT 06269}

\begin{abstract}
The second-order hydrodynamical description of a  homogeneous conformal plasma that undergoes a boost-invariant expansion is given by a single nonlinear ordinary differential equation, whose resurgent asymptotic properties we study, developing further the recent work of Heller and Spalinski [Phys.\ Rev.\ Lett.\  {\bf 115}, 072501 (2015)]. Resurgence clearly identifies the non-hydrodynamic modes that are exponentially suppressed at late times, analogous to the quasi-normal-modes in gravitational language, organizing these modes in terms of a trans-series expansion. These modes are analogs of instantons in semi-classical expansions, where the damping rate plays the role of the instanton action. 
We show that this system displays the generic features of resurgence, with explicit quantitative relations between the fluctuations about different orders of these non-hydrodynamic modes. The imaginary part of the trans-series parameter is identified with the Stokes constant, and the real part with the freedom associated with initial conditions.
\end{abstract}

\maketitle

{\hypersetup{linkcolor=black}
\tableofcontents}

\section{Introduction}
\label{sec:intro}

Resurgent asymptotics using trans-series is a powerful way to extract physical information from asymptotic expansions, systematically including exponentially small contributions that are typically neglected in traditional asymptotic analysis \`a-la Poincar\'e \cite{costin}. This formalism is well developed for ordinary differential equations, both linear and nonlinear, and to a lesser extent some results are known for partial differential equations. Physical applications have included problems in  fluid dynamics \cite{segur}, exact WKB \cite{ZinnJustin:2004ib,delabaere,Dunne:2013ada}, matrix models and strings \cite{Marino:2007te,Pasquetti:2009jg}, and quantum field theory \cite{argyres,cpn,Cherman:2013yfa}. There is also a well-developed literature concerning the 
Painlev\'e transcendents, which themselves have many physical applications, being the nonlinear analogues of the familiar special functions of linear physics \cite{clarkson}. A common thread is the goal to incorporate, in a controlled numerical and analytic manner, exponentially small corrections into asymptotic expansions, in such a way that the trans-series  encodes  its proper analytic continuation properties \cite{Marino:2008ya,Aniceto:2013fka,Dorigoni:2014hea}.

  In this paper we discuss a new example in physics, in the context of hydrodynamics. This is motivated by an interesting recent paper by Heller and Spalinski concerning resummation of the gradient expansion in conformal hydrodynamics \cite{Heller:2015dha}. This example is relevant for hydrodynamic studies of heavy ion collisions \cite{Kolb:2003dz,Teaney:2001av}, and also addresses fundamental issues of the nature of the hydrodynamics expansion. High orders of the gradient expansion of the linearized hydrodynamic equations have been studied in \cite{Lublinsky:2009kv,Grozdanov:2015kqa}. Here we investigate the gradient expansion in the full nonlinear system. 
Our choice of conformal hydrodynamics, and the boost invariant regime thereof, allows us to reduce the hydrodynamic equations from a set of coupled nonlinear partial differential equations to a  single nonlinear {\it ordinary} differential equation. We study in pedagogical detail the asymptotic properties of this nonlinear equation, using resurgent trans-series. We show that this equation displays resurgent trans-series relations between different non-perturbative sectors, and that the asymptotic hydrodynamic expansion encodes detailed quantitative information about non-hydrodynamic modes. These results confirm the relevance of the resurgence approach.
  The ultimate goal is to extend these lessons from the reduced one-dimensional system to the full hydrodynamical system away from the boost-invariant and/or conformal limits, which are presumably non-integrable, and also to the corresponding nonlinear partial differential equations arising through the AdS/CFT correspondence \cite{Horowitz:1999jd,Policastro:2002se,Kovtun:2005ev,Janik:2005zt,Bhattacharyya:2008jc,Kinoshita:2008dq,Janik:2010we}.

\section{Nonlinear equation for Bjorken flow in conformal hydrodynamics}
\label{sec:nonlinear}

In this section we briefly recall the relevant notation for describing second order hydrodynamics and the steps to reduce it to a single nonlinear equation in the restricted case of Bjorken flow in conformal hydrodynamics. The general second order dissipative terms for relativistic, conformal hydrodynamics have been derived in \cite{Baier:2007ix}. The equations of hydrodynamics stem from the conservation of the energy-momentum tensor\footnote{We assume the absence of other conserved charges.}, $\nabla_\mu T^{\mu\nu}=0$. In order to describe the energy-momentum flow we use as our hydrodynamic fields the energy density in the local rest frame of the fluid, $\cal E$, and the fluid four-velocity, $u^\mu$, with $u^\mu u_\mu=-1$. The covariant expression for the energy-momentum tensor in terms of these hydrodynamic fields is
\begin{equation}
T^{\mu\nu}=p(\E) g^{\mu\nu}+\left(p(\E)+ \E\right) u^\mu u^\nu+\Pi^{\mu\nu}
\label{eq:tmunu}
\end{equation}
where $p(\E)$ is the pressure (related to the energy density by the equation of state), and $\Pi^{\mu\nu}$ is the dissipative part that includes the viscous corrections. Conformal symmetry implies $T^\mu_\mu=0$, which determines the equation of the state to be $p(\E)=\E/(d-1)$, for $d$ space-time dimensions. The dissipative part, $\Pi^{\mu\nu}$, is symmetric, transverse (i.e. $u_\mu \Pi^{\mu\nu}=0$), and for conformal fluids, traceless.  In the hydrodynamic limit where one considers only the long wavelength, small momentum modes, the dissipative corrections that constitute $\Pi^{\mu\nu}$ are given as an expansion in space-time gradients of the hydrodynamic fields $\E$ and $u^\mu$ \cite{Baier:2007ix,Romatschke:2009im,Heller:2007qt}. This derivative expansion contains all terms, at a given order, which are allowed by the underlying symmetries of the fluid.  At first order in the derivative expansion, there is only one term allowed by Lorentz and conformal symmetries, the shear viscosity $\eta(\cal E)$:
\begin{equation}
1^{st}\text{ order}:\quad \Pi^{\mu\nu}=-\eta\, \sigma^{\mu\nu}\equiv-2\eta \,^{\langle}\nabla^\mu u^\nu\,^{\rangle} 
\label{eq:pi-1st}
\end{equation}
Here we use the standard notation \cite{Baier:2007ix,Heller:2007qt,Heller:2015dha} to denote the symmetric, transverse projection of a tensor $A^{\alpha\beta}$
\begin{equation}
\,^\langle A^{\mu\nu}\,^\rangle={1\over2}\Delta^{\mu\alpha}\Delta^{\nu\beta}\left(A_{\alpha\beta}+A_{\beta\alpha}\right)-{1\over d-1} \Delta^{\mu\nu} \Delta^{\alpha\beta}A_{\alpha\beta}
\end{equation} 
where $\Delta^{\mu\nu}=g^{\mu\nu}+u^\mu u^\nu$ is the transverse projection operator.

At  second order, there are five new transport coefficients \cite{Baier:2007ix}. One characterizes the relaxation of the energy-momentum tensor, one characterizes the coupling to space-time curvature, and the other three describe non-linear couplings to the fluid velocity. Among these latter three, two are related to couplings to vorticity. In this paper we study homogeneous and boost invariant flow, in which case we can neglect the terms coupling to curvature and vorticity, so only two of the five possible terms contribute:
\begin{equation}
2^{nd}\text{ order}:\quad \Pi^{\mu\nu}=-\eta\, \sigma^{\mu\nu}+ \eta\, \tau_\Pi \left(\,^\langle u^\lambda \nabla_\lambda  \sigma^{\mu\nu}\,^\rangle+{1\over d-1}\sigma^{\mu\nu}\nabla_\lambda u^\lambda\right)+\lambda_1 \sigma^{\langle \mu}_\lambda \sigma^{\nu\rangle \lambda}+\dots
\label{eq:pi-2nd}
\end{equation}
The relevant second order transport coefficients are $\tau_\Pi$ and $\lambda_1$. Using conformal invariance we can parametrize the dimensionful quantites in units of local temperature $T(x)$. For example, the energy density scales as $\E\propto T^d$. For the transport coefficients, we adopt the parameterization used in \cite{Heller:2015dha}
\begin{equation}
 \eta= C_\eta s \quad,\quad \tau_\Pi={C_\tau \over T}\quad,\quad \lambda_1={C_\lambda} {s\over T} 
\label{eq:transport-parametrization}
\end{equation}
where $s\propto T^{d-1}$ is the local entropy density, and $C_\tau$, $C_\lambda$ and $C_\eta$  are dimensionless numbers. 

\subsection{Bjorken flow}
We  consider  boost invariant flow, also known as Bjorken flow \cite{Bjorken:1982qr}. It describes a homogeneous fluid expanding longitudinally along a fixed direction, say $z$. The original motivation of Bjorken flow was to capture the essential physics of the space-time evolution of the quark-gluon plasma produced in heavy ion collisions, which is modeled by a fluid expanding between the highly Lorentz contracted ``sheets" of nuclei, moving away from each other almost at the speed of light, shortly after the collision. With boost invariant initial conditions the expansion is boost invariant, in other words the system looks the same in all inertial frames. It is convenient to work with proper time, $\tau$, and rapidity, $\zeta$ as the space-time coordinates. They are related with the usual Minkowski coordinates $z$ and $t$ as
\begin{equation}
t=\tau\,\cosh\zeta\quad, \quad z=\tau\,\sinh\zeta\quad\Leftrightarrow\quad \tau=\sqrt{t^2-z^2}\quad,\quad \zeta=\tanh^{-1}(t/z)\,.
\end{equation}
The metric is $ds^2=-d\tau^2+\tau^2d\zeta^2+dx^2_{\perp}$ and the non-vanishing Christoffel symbols are $\Gamma^\zeta_{\tau\zeta}=\tau^{-1}$, and $\Gamma^\tau_{\zeta\zeta}=\tau$. In these coordinates the fluid velocity is constant, $u^\tau=1$, $u^\xi=u^{x_\perp}=0$, and the conservation equation, $\nabla_\mu T^{\mu\nu}=0$, along with Eq. \eqref{eq:tmunu} and the conformal equation of state, $p=\E/(d-1)$, reduce to
\begin{equation}
\tau \frac{d \E}{d\tau}+\frac{d}{d-1} \E- \Phi=0 
\label{eq:epsilon-eqn}
\end{equation}
where $\Phi\equiv-\Pi^\zeta_\zeta$ \cite{Baier:2007ix}. At leading order where the viscous terms $\Phi$ are neglected, the energy density has the asymptotic large $\tau$ behavior 
\begin{equation}
\E\sim\tau^{-{d\over d-1}}+\dots \quad, \quad \left(\E\sim\tau^{-{4/3}}+\dots \quad , d=4\right)
\label{eq:epsilon-leading}
\end{equation}
Here the ellipses denote the viscous corrections. Equation (\ref{eq:epsilon-leading}) is the leading order result that describes how the system cools as it expands. The viscous corrections can be calculated by inserting \eqref{eq:pi-2nd} into \eqref{eq:epsilon-eqn} and solving as an expansion at  large proper-time $\tau$. We instead follow a M\"uller-Israel-Stewart-like approach and promote the dissipative part of the energy-momentum tensor $\Pi^{\mu\nu}$ to an independent hydrodynamic field.\footnote{The M\"uller-Israel-Stewart approach  \cite{mis}  was designed to obtain a set of hyperbolic differential equations which have causal solutions as opposed to the hydrodynamical equations which are not hyperbolic and have acausal propagation of certain modes. These modes are beyond hydrodynamics, but eliminating them has advantages in numerical simulations.} Using the first order relation \eqref{eq:pi-1st} we can rewrite \eqref{eq:pi-2nd}, also to second order, as
 \begin{equation}
 \Pi^{\mu\nu}=-\eta\, \sigma^{\mu\nu}- \eta\, \tau_\Pi \left(\,^\langle u^\lambda \nabla_\lambda  \Pi^{\mu\nu}\,^\rangle+{1\over d-1}\Pi^{\mu\nu}\nabla_\lambda u^\lambda\right)+\lambda_1 \Pi^{\langle \mu}_\lambda \Pi^{\nu\rangle \lambda}
 \label{eq:pi-ims}
 \end{equation}
and obtain a relaxation equation for $\Pi^{\mu\nu}$, with  relaxation time $\tau_\Pi$. This procedure generates an all-orders derivative expansion which agrees with the ordinary hydrodynamic expansion up to third order, beyond which more transport coefficients should be taken into account. 

In the homogeneous, boost invariant limit the relaxation equation \eqref{eq:pi-ims} further reduces to a non-linear equation
 \begin{equation}
 \tau_\Pi\, \frac{d \Phi}{d\tau}+\left(1+\frac{d}{d-1}\frac{\tau_\Pi}{\tau}\right)\, \Phi+\left(\frac{d-3}{d-2}\right) \frac{\lambda_1}{\eta^2} \Phi^2 -2\left(\frac{d-2}{d-1}\right)\frac{\eta}{\tau}=0
 \label{eq:phi-eqn}
 \end{equation} 
 where the transport coefficients $\eta$, $\tau_\Pi$ and $\lambda_1$ are functions of the energy density $\E$. The technical problem  now is to solve the two coupled nonlinear equations \eqref{eq:epsilon-eqn} and \eqref{eq:phi-eqn}. 

To illustrate the ideas of resurgence in hydrodynamics, we consider as our central object the all-orders expansion generated by this M\"uller-Israel-Stewart-like treatment of the second order, conformal, boost invariant hydrodynamics. In $d=4$ the resummation of this expansion is studied by Heller and Spalinski in \cite{Heller:2015dha}. We show that the usual derivative expansion encodes much more information, such as the non-hydrodynamical modes, than one might naively anticipate, a result which  we argue has implications for more complicated and phenomenologically relevant models as well, such as those relating hydrodynamics to gravitational systems \cite{Policastro:2002se,Kovtun:2005ev,Bhattacharyya:2008jc,Janik:2010we}.

\subsection{Scale invariance and the nonlinear hydrodynamic equation}

Using the parametrization of the transport coefficients \eqref{eq:transport-parametrization} dictated by conformal invariance, equations \eqref{eq:epsilon-eqn} and \eqref{eq:phi-eqn} combine into a single highly nonlinear equation for the temperature $T(\tau)$:
\begin{eqnarray}
&& C_\tau {\tau^2 \ddot T\over T}+{ (d-3)(d-1)C_\lambda\over (d-2)C_\eta}{\tau^3 \dot T^2\over T}+ (d-1)C_\tau\, {\tau^2\dot T^2\over T^2}+{ (3d-1)C_\tau\over d-1} {\tau \dot T\over T}+\left(1+{2(d-3)C_\lambda\over (d-2)C_\eta}\right)\tau^2 \dot T\nonumber\\
&&+{1\over d-1}\left(1+{(d-3)C_\lambda\over (d-2)C_\eta}\right) \tau T+{d\,C_\tau-2(d-2) C_\eta\over(d-1)^2}=0
\label{eq:T-eqn}
\end{eqnarray}
where $\dot T\equiv dT/d\tau$. As a consequence of the underlying scale invariance of the system, Eq. \eqref{eq:T-eqn} is invariant under rescaling $\tau\rightarrow\alpha\tau$, $T\rightarrow\alpha^{-1}T$. Therefore we can integrate Eq. \eqref{eq:T-eqn} to obtain a first order equation. Scale invariance also suggests introducing a local proper-time variable, $w$, measured in units of local temperature (in mathematical terms, $w$ is known as the ``\'Ecalle time''):
\begin{equation}
w\equiv \tau \, T(\tau)
\label{eq:w-defn}
\end{equation}
and an associated dimensionless characterization of the temperature, through a  function $f(w)$ defined as\footnote{Note that our $f$ differs from the $f$ in \cite{Heller:2015dha} by: $f_{\rm ours}=f_{\rm theirs}-1$.}  
\begin{equation}
f\equiv {d\log T\over d\log\tau}
\label{eq:f-defn}
\end{equation}
With these definitions,  Eq. \eqref{eq:T-eqn} reduces to a single first order nonlinear equation for the dimensionless variable $f(w)$ as a function of $w$, the proper time measured in the units of local temperature. 
\begin{eqnarray}
&&C_\tau w f(w) f^\prime(w)+C_\tau w f^\prime(w)+\left(1+{2(d-3)C_\lambda\over (d-2)C_\eta}\right) w f(w)+ d\,C_\tau f^2(w)+{ (d-3)(d-1) C_\lambda\over (d-2)C_\eta} wf^2(w)
\nonumber\\
&&+{1\over d-1}\left(1+{(d-3)C_\lambda\over (d-2)C_\eta}\right)w+{2d\,C_\tau\over d-1}f(w)+{d\,C_\tau-2(d-2) C_\eta\over(d-1)^2}=0
\label{eq:f-ddim-eqn}
\end{eqnarray}
(Here we study this equation in $d=4$, but we note that $d=3$ also has novel features.)
For $d=4$, we reach the central equation to be studied in this paper:
\begin{eqnarray}
&&C_\tau w f(w) f^\prime(w)+C_\tau w f^\prime(w)+\left(1+{C_\lambda\over C_\eta}\right) w f(w)+4 C_\tau f^2(w)+{3 \,C_\lambda \over 2 \,C_\eta} wf^2(w)
\nonumber\\
&&+{1\over 3}\left(1+{C_\lambda\over 2 \,C_\eta}\right)w+{8 \,C_\tau\over 3}f(w)+{4\over9}{(C_\tau-C_\eta)}=0
\label{eq:main-eqn}
\end{eqnarray}
Up to the simple shift,  $f_{\rm ours}=(f_{\rm theirs}-1)$ noted above, (\ref{eq:main-eqn}) is the equation analyzed in \cite{Heller:2015dha}.

It is clear that after the rescaling $w\rightarrow C_\tau w$, Eq. \eqref{eq:main-eqn} only depends on the ratios $C_\eta/C_\lambda$, and $C_\eta/C_\tau$. Physically, this is simply because $C_\tau$ characterizes the relaxation time, using $w$ now as the time variable, and so is naturally absorbed into $w$. However, to facilitate direct comparison with previous results \cite{Heller:2015dha}, we will not adopt this rescaling. And in what follows, for concrete numerical illustrations, we will also use the parameters that are associated with ${\cal N}=4$ super Yang-Mills theory \cite{Baier:2007ix,Heller:2007qt,Heller:2015dha}:
\begin{equation}
C_\eta={1\over 4\pi}\quad,\quad C_\tau={2-\log2\over 2\pi}\quad,\quad C_\lambda={1\over 2\pi}
\label{eq:N4}
\end{equation}

\section{Formal late time expansions}
\label{sec:formal}

In this section we analyze the hydrodynamic expansion generated by the nonlinear equation Eq. \eqref{eq:main-eqn}. The hydrodynamic regime is identified with large proper time, or equivalently large $w$.  Mathematically speaking the large $w$ expansion is a formal series which is divergent, asymptotic and non-Borel summable. Such asymptotic behavior is characteristic of gradient expansions of effective actions \cite{Dunne:1999uy}. This is by no means a bad thing: the asymptotic nature of the expansion actually encodes important physical information.

\subsection{Late time hydrodynamic expansions}
\label{sec:late_time}

The late proper-time hydrodynamic expansion\footnote{As terminology, we refer to the all-orders expansion \eqref{eq:f0-ansatz} generated by the M\"uller-Israel-Stewart-type analysis described in Section \ref{sec:nonlinear} as the ``hydrodynamic expansion''. We will see that a consistent asymptotic analysis of the basic nonlinear equation \eqref{eq:main-eqn} requires the addition of further terms, which go beyond hydrodynamics.} of $f(w)$ is an expansion in inverse powers of $w$. Since $w\sim \tau^{2/3}$ in $d=4$, this expansion translates into a late proper-time expansion for $T(\tau)$ using (\ref{eq:w-defn}, \ref{eq:f-defn}). We start with the formal series ansatz
\begin{equation}
f^{(0)}(w)=\sum_{k=0}^\infty f^{(0)}_k\, w^{-k}\,.
\label{eq:f0-ansatz}
\end{equation}
The meaning of the superscript $(0)$ will become clear in Section \ref{sec:trans-series}, when we include the terms beyond hydrodynamics. With this ansatz \eqref{eq:f0-ansatz}, the nonlinear equation \eqref{eq:main-eqn} generates a recursion relation for the the coefficients $f^{(0)}_k$:
\begin{eqnarray}
&&-C_\tau  \sum_{k'=0}^{k}\,(k' - 1)\,f^{(0)}_{k-k'}\,f^{(0)}_{k'-1}-C_\tau (k - 1)\,f^{(0)}_{k-1}+\left(1+{ C_\lambda \over C_\eta}\right)f^{(0)}_{k}+4\, C_\tau  \sum_{k'=0}^{k-1} f^{(0)}_{k-k'-1}f^{(0)}_{k'}\nn
&&+{3\, C_\lambda \over2\, C_\eta} \sum_{k'=0}^{k} f^{(0)}_{k-k'}f^{(0)}_{k'} +{1 \over3} \left( 1+{C_\lambda \over 2\,C_\eta}\right)\delta_{k,0}+{8\,C_\tau\over3} f^{(0)}_{k-1}+{4\over9}\left( C_\tau- C_\eta\right)\delta_{k-1,0}=0
\label{eq:rec-pert}
\end{eqnarray}
These can be solved iteratively. The equation that determines $f^{(0)}_0$ is quadratic. Therefore there are actually two solutions to the recursion relations,  leading to two different large $w$ expansions:
\begin{eqnarray}
f^{(0)}_+(w)&\sim& -\frac{1}{3} +\frac{4 C_\eta}{9} \frac{1}{w} -\frac{8 C_\eta (C_\lambda-C_\tau)}{27}\frac{1}{w^2} +{16\,C_\eta\, (2(C_\lambda-C_\tau)^2-3C_\eta\,C_\tau)\over 81 }{1\over w^3}+\dots 
\label{eq:late1}
\\
f^{(0)}_-(w)&\sim& -\frac{1}{3}\left(1+\frac{2\,C_\eta}{C_\lambda}\right) -\frac{4 C_\eta}{9} \left(1-\frac{4C_\eta\, C_\tau}{C_\lambda^2}\right) \frac{1}{w}+ \dots 
\label{eq:late2}
\end{eqnarray}
In terms of physical quantities, such as the proper time $\tau$ and the local temperature $T(\tau)$, it is the former expansion, $f^{(0)}_+$, which leads to the familiar hydrodynamic expansion $T(\tau)\sim \tau^{-1/3}-{2 C\eta \over 3 \tau}+\dots$. On the other hand,  the latter expansion $f^{(0)}_-$ leads to  $T(\tau)\sim \tau^{-\frac{1}{3}\left(1+\frac{2\,C_\eta}{C_\lambda}\right) }+\dots$, which does not connect to the ideal hydrodynamic result at late times. Furthermore, the expansion $f^{(0)}_-(w)$ in (\ref{eq:late2}) is unstable, in a sense explained below, so we concentrate on the expansion $f^{(0)}_+$ from now on. With this understood, for notational simplicity we drop the subscript $``+"$, and simply write $f^{(0)}(w)$ for the function with late time expansion in (\ref{eq:late1}).

\subsection{Divergence, Borel ambiguities, and exponentially suppressed terms}
\label{sec:divergence-pert}

It is straightforward to show that the expansion \eqref{eq:late1} is divergent. The late terms in the expansion grow factorially fast, and their leading large-order growth can be characterized as
\begin{equation}
f^{(0)}_{k}\sim \frac{\Gamma(k+\beta)}{S^{k+\beta}} \quad, \quad k\to \infty
\label{eq:large-order-leading}
\end{equation}
for some real numbers $S$ and $\beta$, whose physical meaning will become clear shortly. Furthermore, with the ${\mathcal N}=4$ parameters in (\ref{eq:N4}), the parameter $S$ turns out to be positive. There are three important and interrelated points related to this divergent behavior  \cite{bender,costin}. First,  the hydrodynamic expansion (\ref{eq:f0-ansatz}) is a typical asymptotic expansion: for a fixed large value of $w$, the terms $f^{(0)}_k w^{-k}$ at first decrease for increasing low values of $k$, but eventually start growing beyond some value of $k$, say $k^*$. This transition happens roughly when $\left.d (f^{(0)}_k w^{-k})/dk \right |_{k^*}=0$, or $k^*\approx S w$. One can obtain exponential precision by the procedure of ``least-term-truncation'', truncating the series at $k\sim k^*$. The associated inherent error is exponentially small
\begin{equation}
f^{(0)}_{k^*}w^{-k^*}\sim w^{\beta}\,e^{-Sw} 
\label{eq:error}
\end{equation}
Second, one could try to apply Borel summation:
\begin{eqnarray}
f^{(0)}(w)\sim\sum_{n=0}^\infty {\Gamma(n+\beta) \over S^{n+\beta} w^n}&=& \sum_{n=0}^\infty w^\beta   \int_0^\infty \frac{du}{u} \, e^{-S w u}\,u^{n+\beta}  \sim \int_0^\infty \frac{du}{u} \, e^{-S w u}\,{u^{\beta}\over 1-u } \,.
\end{eqnarray}
However, when $S>0$, as is the case here, the Borel integral has a pole at $u=1$,  leading to an ambiguous imaginary part 
\begin{equation}
{\cal I}m [f^{(0)}(w)]\sim \mp i \pi\, w^{\beta}\,e^{-Sw}
\label{eq:ambig}
\end{equation}
Notice that this is the same order of magnitude as the least-term-truncation error in (\ref{eq:error}).
 The ambiguity (\ref{eq:ambig}) is due to the ambiguity of deforming the integration contour around the pole. The existence of such an ambiguous term is problematic for several reasons: it has an undetermined sign, and furthermore it is pure imaginary. However the function $f(w)$ should be real, since it is related to the physical temperature. It is clear that in order to fix this ambiguity we need to add to $f^{(0)}(w)$ an exponentially suppressed term of the same order.

Third, one can consider linearized perturbations around the solution $f^{(0)}(w)$. This is achieved by an ansatz $f^{(0)}(w)\rightarrow f^{(0)}(w)+\delta f^{(0)}(w)$. Inserting this ansatz into Eq. \eqref{eq:main-eqn}, keeping the first two terms in the $w^{-1}$ expansion of $f^{(0)}(w)$ in \eqref{eq:late1} and the linear terms in $\delta f^{(0)}(w)$ leads to the equation
\begin{equation}
0=w\left({2 C_\tau\over 3}{d (\delta f^{(0)})\over dw}+\delta f^{(0)}\right)+{4 C_\eta C_\tau \over 9}{d (\delta f^{(0)})\over dw}+{4C_\lambda\over 3}\delta f^{(0)}+{\cal O}(w^{-1})
\label{eq:lin-pert-eqn}
\end{equation}
which has the solution\footnote{The same argument for the other expansion $f^{(0)}_-$, in Eq. \eqref{eq:late2}, leads to 
$\delta f^{(0)}_-(w) \sim w^{\beta}\, \exp\left[-\frac{3 C_\lambda}{2 C_\tau(C_\eta-C_\lambda)} \,w\right]$ with $\beta=\frac{2C_\lambda^3-C_\eta C_\lambda^2-4C_\eta^2 C_\tau}{C_\tau (C_\eta-C_\lambda)^2}$. Notice that for the ${\cal N}=4$ parameters \eqref{eq:N4}, the exponent is positive, which is related to the unstable nature of the expansion in \eqref{eq:late2}.}
\begin{eqnarray}
\delta f^{(0)}(w) &\sim& w^{\beta}\, \exp\left[-\frac{3}{2 C_\tau} \, w\right]\quad, \quad \beta=\frac{C_\eta-2C_\lambda}{C_\tau}
\label{eq:lin-pert}
\end{eqnarray}
Notice that the linearized perturbation in (\ref{eq:lin-pert}) has the same functional form as the least-term-truncation error (\ref{eq:error}), and the ambiguity (\ref{eq:ambig}) of naive Borel summation. 
This is not a coincidence, as all  three phenomena are manifestations of the asymptotic character  of the hydrodynamic expansion (\ref{eq:f0-ansatz}). In fact, by general arguments \cite{costin,bender} the constants $S$ and $\beta$ appearing in the leading large order growth (\ref{eq:large-order-leading}), and hence also in the error  (\ref{eq:error}) and the ambiguity (\ref{eq:ambig}), are exactly the same constants that appear in the linearized perturbation (\ref{eq:lin-pert}). Thus we deduce that
\begin{equation}
S={3\over2C_\tau}\quad,\quad\beta=\frac{C_\eta-2C_\lambda}{C_\tau}
\label{eq:sbeta}
\end{equation}
which becomes now a  {\it numerical prediction} for the leading large-order growth (\ref{eq:large-order-leading}) of the coefficients of the hydrodynamic expansion, which are generated from the recursion relations (\ref{eq:rec-pert}). Numerically, for the  ${\cal N}=4$ SYM plasma parameters in (\ref{eq:N4}):
 \begin{equation}
S={3 \pi \over 2-\log2}\approx7.21181\quad,\quad\beta=-{3\over 4-2\log2}\approx-1.1478
\end{equation}
In Section \ref{sec:large_order} we confirm the consistency of these arguments with great precision.

The conclusion is that by itself the hydrodynamic series expansion $f^{(0)}(w)$ in (\ref{eq:f0-ansatz}) is merely a formal expression. In order to promote it to a well-defined function it is necessary to enhance it by adding some exponential terms that go beyond the hydrodynamic expansion. These terms are not arbitrary: they are highly constrained. The linearized perturbation is only the first element of an infinite set of exponential corrections. These exponential terms are required to render an unambiguous answer for the actual solution, $f(w)$, of the differential equation. The full expansion that contains all these exponential terms in addition to the formal series expansion (\ref{eq:f0-ansatz})  is known as a ``trans-series'' \cite{costin}. In the next section we demonstrate how such a trans-series is constructed.

\section{Trans-series expansions}
\label{sec:trans-series}

The leading exponential correction to the perturbative series, $\delta f^{(0)}(w)$,  is only the tip of the iceberg. Just like the hydrodynamic series $f^{(0)}(w)$ itself, the first exponential correction is also an asymptotic series: the factor $w^\beta\,e^{-S w}$  is multiplied by another formal series in $w^{-1}$, which we denote as $f^{(1)}(w)$, and which is also divergent, and with the same leading rate of growth (\ref{eq:large-order-leading}) as in the hydrodynamic series. By the same arguments, we conclude that there must be a correction of the form $w^{2\beta}\,e^{-2Sw}$. Furthermore, this second exponential term is also multiplied by another asymptotic series, denoted by $f^{(2)}(w)$, which requires the existence of corrections of the from $w^{3\beta}\, e^{-3Sw}$. This pattern continues ad infinitum. (This is also clear generically from the nonlinearity of the underlying differential equation.)  In order to obtain an unambiguious answer, \textit{all} these exponential terms, and their fluctuations,  must be included in the answer. The resulting combined object is known as a ``trans-series''.\footnote{More general trans-series also include powers of logarithms, and possibly iterations of powers, exponentials and logarithms \cite{costin,Aniceto:2013fka,Dorigoni:2014hea}.} 
More importantly, the theory of resurgence predicts that these various asymptotic series, which make up the full trans-series, are related to one another in extremely intricate ways. These inter-relations will be demonstrated explicitly in Section \ref{sec:large_order}.

We are led to the following trans-series ansatz generalization of the formal hydrodynamic expansion  (\ref{eq:f0-ansatz}):
\begin{eqnarray}
f(w)&\sim&  f^{(0)}(w) + \sigma\, w^\beta \,e^{-S w } \,f^{(1)}(w) + \sigma^2\, w^{2\beta} \,e^{-2 S w } \,f^{(2)}(w) +\dots 
\label{eq:trans-ansatz}\\
&\sim & \sum_{n=0}^\infty f^{(n)}(w)\, \sigma^n \, \zeta^n(w) 
\label{eq:trans2}
\end{eqnarray}
This is a sum over powers of an exponential factor (the analog of the ``instanton fugacity" in semi-classical expansions in quantum mechanics or quantum field theory, with $S$ being the instanton action)
\begin{eqnarray}
\zeta(w)\equiv w^\beta \,e^{-S w }
\label{eq:xi}
\end{eqnarray}
each multiplied by a formal large $w$ series (the analog of the pertrurbative fluctuations about the $n^{\rm th}$ instanton sector):
\begin{eqnarray}
f^{(n)} (w)\sim f^{(n)}_0+\frac{1}{w} f^{(n)}_1+\frac{1}{w^2} f^{(n)}_2+\frac{1}{w^3} f^{(n)}_3 +\dots
\label{eq:trans3}
\end{eqnarray}
In the trans-series expansion (\ref{eq:trans-ansatz}, \ref{eq:trans2}),  $\sigma$ is the trans-series expansion parameter, which is in general a complex number. As discussed below, its imaginary part will be fixed by resurgent cancellations associated with the reality of the trans-series, while its real part is a free parameter related to the initial conditions of the ODE.  At this stage, it simply counts the ``non-perturbative'' order of the trans-series expansion. 

Inserting the trans-series ansatz \eqref{eq:trans-ansatz} into the nonlinear ODE \eqref{eq:main-eqn}, we obtain a set of recursion relations by equating to zero the coefficient of each $w^{n\beta-k}e^{-n S w}$, for all $k\geq0$ and $n\geq0$. Setting $n=0$ leads to the recursion relations \eqref{eq:rec-pert} that generate the hydrodynamic series \eqref{eq:late1}  studied in the previous section. Setting $n=1$, we obtain the recursion relations that generate the asymptotic series $f^{(1)}(w)$ that multiplies the first exponentially suppressed term, or the first non-hydrodynamic series. We refer to the asymptotic series $f^{(n)}(w)$ as the  ``$n^{th}$ non-hydrodynamic series".
\bear
&&-C_\tau \sum_{k'=0}^{k} \left[ f^{(1)}_{k-k'}\, f^{(0)}_{ k' - 1} \left( k' - 1\right )+ f^{(0)}_{k-k'} \left(  S\,  f^{(1)}_{k'} +(k' - \beta- 1)\, f^{(1)}_{ k' - 1}\right)\right]+\left(1+{ C_\lambda \over C_\eta}\right) f^{(1)}_{k}
\nn
&&-C_\tau  \left(  S\,  f^{(1)}_{k} +(k -  \beta- 1)\, f^{(1)}_{ k - 1}\right)+4\, C_\tau  \sum_{k'=0}^{k-1}\left[  f^{(1)}_{k-k'-1} f^{(0)}_{k'}+ f^{(0)}_{k-k'-1} f^{(1)}_{k'}\right]
\nn
&&+{3\, C_\lambda \over2\, C_\eta} \sum_{k'=0}^{k} \left[ f^{(1)}_{k-k'} f^{(0)}_{k'}+ f^{(0)}_{k-k'} f^{(1)}_{k'} \right]+{8\,C_\tau\over3}   f^{(1)}_{k-1}=0\,.
\label{eq:rec-1}
\eear
The first two terms, $k=0$ and $k=1$, in these recursion relation determine the constants $S$ and $\beta$ to be as in (\ref{eq:lin-pert}). The leading coefficient $f^{(1)}_0$ is a free parameter, and the remaining coefficients, $f^{(1)}_k$ ($k\geq 1$), are determined by Eq. \eqref{eq:rec-1}, and are all proportional to $f^{(1)}_0$. Therefore, without  loss of generality we can normalize $f^{(1)}_0=1$, and characterize the freedom in terms of the constant $\sigma$.  This normalization fixes all the other coefficients $f^{(1)}_{k}$ uniquely. 
The leading non-hydrodynamic series is 
\bear
f^{(1)}(w)\sim 1+ \frac{2  \left(C_\eta^2-C_\eta (C_\lambda-6 C_\tau)+2 C_\lambda (C_\tau-C_\lambda)\right)}{3 C_\tau\, w}+\dots
\label{eq:f1}
\eear

The higher hydrodynamic series are determined by similar recursion relations. For completeness we give the full recursion relation for arbitrary $n$ and $k$:
\bear
&&-C_\tau \sum_{n'=0}^n \sum_{k'=0}^{k} f^{(n-n')}_{k-k'} \left( n'\, S\,  f^{(n')}_{k'} +(k' - n' \beta- 1)\, f^{(n')}_{ k' - 1}\right)-C_\tau  \left( n S\,  f^{(n)}_{k} +(k - n \beta- 1)\, f^{(n)}_{ k - 1}\right)
\nn
&&+\left(1+{ C_\lambda \over C_\eta}\right) f^{(n)}_{k}+4\, C_\tau \sum_{n'=0}^n \sum_{k'=0}^{k-1}  f^{(n-n')}_{k-k'-1} f^{(n')}_{k'}+{3\, C_\lambda \over2\, C_\eta}\sum_{n'=0}^n \sum_{k'=0}^{k}  f^{(n-n')}_{k-k'} f^{(n')}_{k'}+{8\,C_\tau\over3}  f^{(n)}_{k-1}
\nn
&&+{1 \over3} \left( 1+{C_\lambda \over 2\,C_\eta}\right)\delta_{k,0}\delta_{n,0}+{4\over9}\left( C_\tau- C_\eta\right)\delta_{k-1,0}\delta_{n,0}=0
\label{eq:rec-n}
\eear
Once the free parameter $f^{(1)}_0$, or $\sigma$, is fixed there is no more freedom, and all the coefficients $f^{(n)}_k$ are determined completely by the recursion relations \eqref{eq:rec-n}. This is perhaps surprising, but it is generic \cite{costin}.
The dependence of the trans-series on $\sigma$ is exactly as written in Eq. \eqref{eq:trans2}, i.e. $\sigma$ enters the trans-series as $\sigma^n f^{(n)}(w)$.  For example, the second non-hydrodynamic expansion is
\bear
f^{(2)}(w)&\sim&{3 (C_\lambda-C_\eta)\over2C_\eta}+ \frac{ -2 C_\eta^3+C_\eta^2 (4 C_\lambda-11 C_\tau)+2 C_\eta\left(C_\lambda^2+4 C_\lambda C_\tau+2 C_\tau^2\right)+2 C_\lambda^2 (C_\tau-2 C_\lambda)}{C_\eta C_\tau w}
\nn
&&+\dots
\label{eq:f2}
\eear
and the  third non-hydrodynamic series is 
\bear
f^{(3)}(w)&\sim& \frac{9 (3 C_\eta-2 C_\lambda) (C_\eta-C_\lambda)}{8 C_\eta^2}+\frac{1}{4 C_\eta^2 C_\tau w}\left[C_\tau \left(48 C_\eta^3-67 C_\eta^2 C_\lambda+16 C_\eta C_\lambda^2+4 C_\lambda^3\right)+\right.
\nn 
&&\left. 4 C_\eta C_\tau^2 (4 C_\lambda-5 C_\eta)+3 (C_\eta-C_\lambda) (C_\eta+C_\lambda) (C_\eta-2 C_\lambda) (3 C_\eta-2 C_\lambda)\right]
+\dots
\label{eq:f3}
\eear
The remaining coefficients can be generated in a straightforward fashion, but are rather cumbersome to write.

More physically, these non-hydrodynamical series correspond to exponentially damped modes multiplied by a gradient expansion due to viscous terms. Recalling $w\sim \tau^{2/3}$, these modes contribute to expansions of physical quantities such as the local temperature, or energy density as 
\begin{equation}
\zeta^n(w) \sim \tau^{ {2n\beta\over3}}e^{-n{3\over 2 C_\tau}\tau^{2/3}} \sim\tau^{2n\beta\over3}e^{-n{3\over 2 T_0}{\tau\over\tau_\Pi}}\,.
\end{equation}
Note that for the ${\cal N}=4$ parameters (\ref{eq:N4}) these modes introduce a transcendental power (i.e. $\beta$) of $\tau$ in the gradient expansion. It is illustrative to compare the non-hydrodynamic modes to instanton contributions in quantum mechanics and quantum field theory. The coefficient $S$ that controls the damping is analogous to the two-instanton action in the dilute instanton gas picture, where the ambiguities that arise from the divergence of perturbation theory are cured by non-perturbative corrections from higher (multi-)instanton sectors \cite{ZinnJustin:2004ib,cpn,Dunne:2013ada}. Here the non-hydrodynamic modes play the role of instantons. 

\section{Large order behavior, Stokes constants and Borel analysis}
\label{sec:large_order}

The {\it reality} condition on the trans-series $f(w)$ in (\ref{eq:trans-ansatz}, \ref{eq:trans2}) means that cancellations must occur of imaginary terms generated by Borel summation of the different $f^{(n)}(w)$ expansions. As described in the comprehensive analysis of real trans-series by Aniceto and Schiappa \cite{Aniceto:2013fka}, this leads to an infinite hierarchy of relations between the expansion coefficients $f_k^{(n)}$ of different $n$ sectors. In this Section we analyze these relations numerically, and confirm  that these  resurgence relations do indeed hold. This also allows us to deduce the Borel plane structure of the associated Borel transforms, which reveals some interesting branch-cut structure.

\subsection{Hydrodynamic derivative expansion: $f^{(0)}(w)$}
\label{sec:zero_expansion}

Resurgence predicts \cite{Aniceto:2013fka} that the leading large-order growth of the expansion coefficients $f^{(0)}_k$ in (\ref{eq:large-order-leading}) receives subleading corrections of the form:
\bear
f^{(0)}_k&\sim& S_1\,{\Gamma(k+\beta)\over 2 \pi i \,S^{k+\beta}}\left(f^{(1)}_0+{S\over k+\beta-1}\,f^{(1)}_1+{S^2\over (k+\beta-1)(k+\beta-2)}\,f^{(1)}_2+\dots\right)+\dots\,.
\label{eq:zero-growth}
\eear
This is an example of the resurgent relations between two sectors of the trans-series, namely between  the hydrodynamic $f^{(0)}(w)$ and the first non hydrodynamic series $f^{(1)}(w)$. Recall that the constants $S$ and $\beta$ appearing in (\ref{eq:zero-growth}) are known, being given by  (\ref{eq:sbeta}). The subleading terms are also known, as they are completely determined by the low order coefficients, $f_0^{(1)}$, $f_1^{(1)}$, $f_2^{(1)}$, $\dots$,  of the first non-hydrodynamic series $f^{(1)}(w)$ in (\ref{eq:f1}). It is therefore possible to determine numerically the overall constant  $S_1$, a ``Stokes constant''\footnote{The factor $2\pi i$ is the denominator is conventional, being convenient for the associated Borel analysis.},  by generating many coefficients  $f^{(0)}_k$ from the recursion relations (\ref{eq:rec-pert}). It is simple to generate many thousands of these coefficients, which provides great precision.

\begin{figure}[h]
\includegraphics[width=0.495\textwidth]{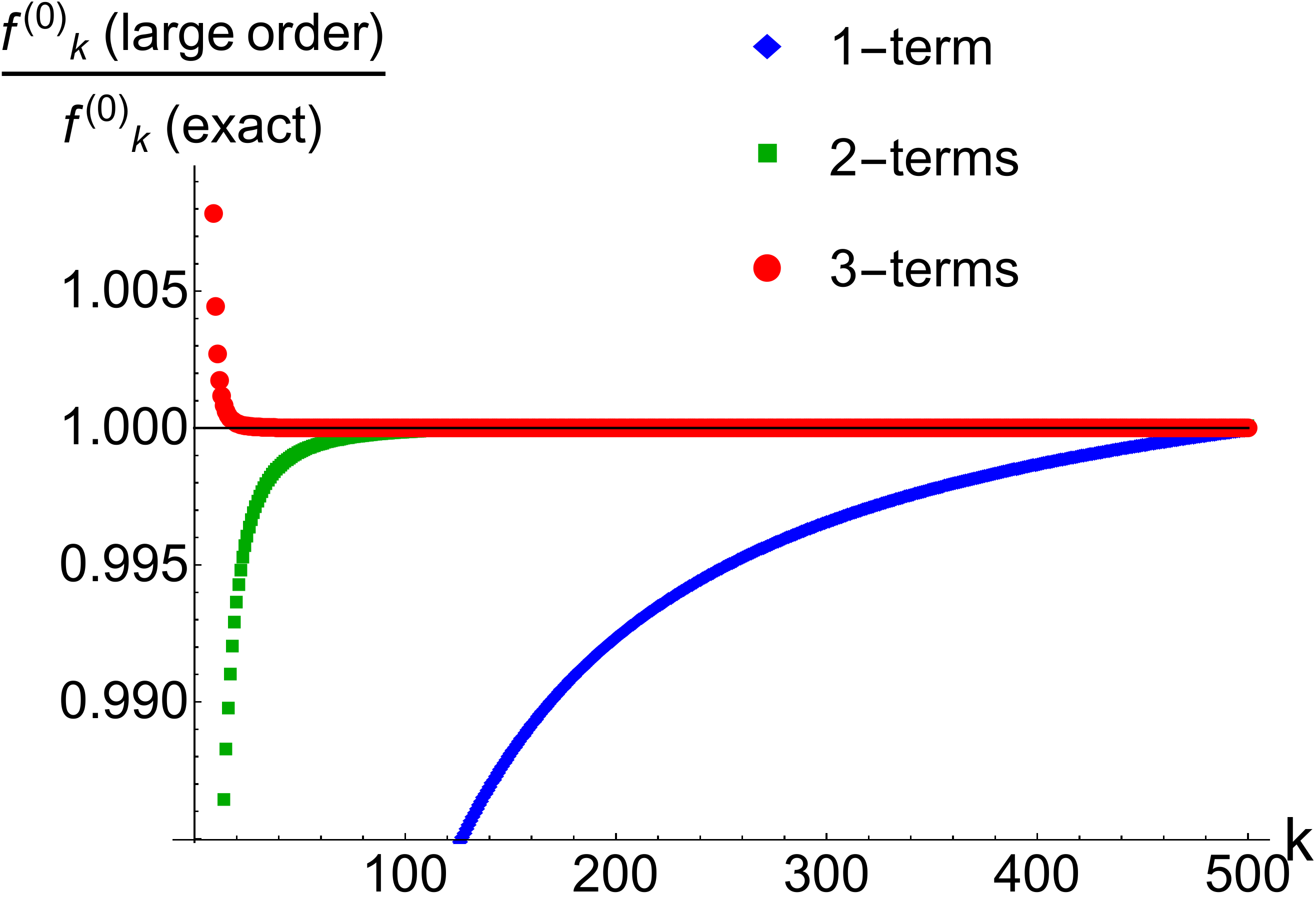}
\includegraphics[width=0.495\textwidth]{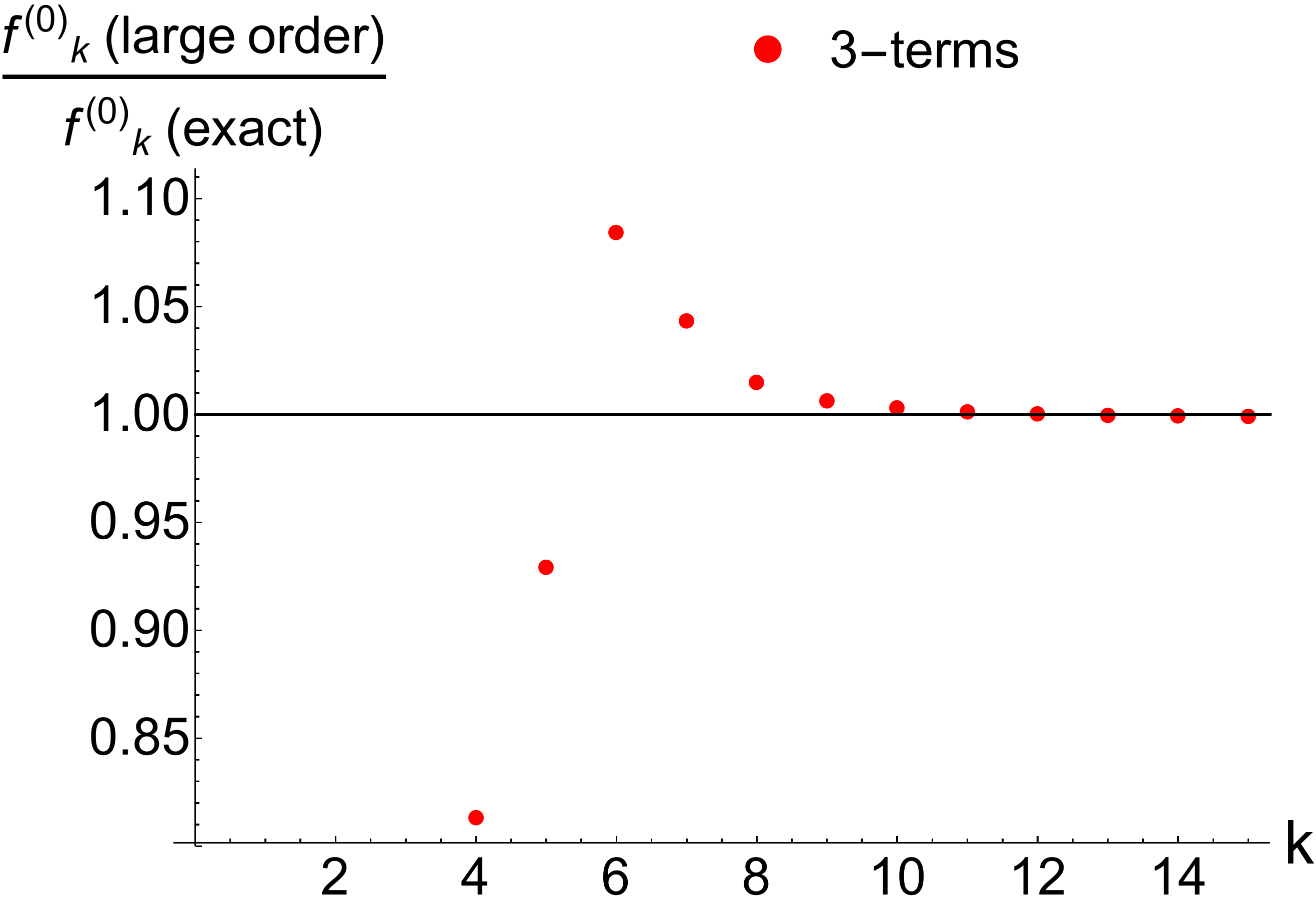}
\caption{The first plot shows plot the ratio of the large order expression \eqref{eq:zero-growth} to the exact coefficients $f^{(0)}_k$, as a function of $k$. The blue, green and red points refer to the inclusion of the zeroth, first, and second subleading terms in \eqref{eq:zero-growth}. The second plot shows a close-up view of the ratio using the  expression with the first three terms in \eqref{eq:zero-growth}. This clearly demonstrates the remarkable precision of the resurgence relation \eqref{eq:zero-growth}:  the agreement is better than the one percent level already by $k\approx 10$.}
\label{fig:0-lo}
\end{figure}

 In Figure \ref{fig:0-lo} we plot the ratio of the large order expression \eqref{eq:zero-growth} to the actual coefficients. This clearly demonstrates the remarkable precision of the resurgence relation \eqref{eq:zero-growth}. Indeed, as shown in the second plot, when all three terms on the right-hand-side of \eqref{eq:zero-growth} are included, the agreement with the prediction of the large-order growth is at the one percent level already by the $10^{\rm th}$ term in the expansion.
From direct comparison of the large order expression \eqref{eq:zero-growth} with the exact coefficients $f^{(0)}_k$ we  deduce the numerical value of the Stokes constant $S_1$:
\begin{equation}
S_1\approx-0.040883\, i \qquad, \quad  \text{for ${\cal N}=4$ parameters}
\label{eq:stokes}
\end{equation}
Note that $S_1$ is pure imaginary.

The form of the large order growth is intimately connected with the Borel transform of the asymptotic series $f^{(0)}(w)$. The Borel transform is defined as
\bear
{\hat f^{(0)}}(s)\equiv\sum_{k=0}^\infty {f^{(0)}_{k+1}\over k!} s^k \,.
\label{eq:borel-def}
\eear
For factorially divergent series such as $f^{(0)}(w)$, this new function ${\hat f^{(0)}}(s)$ has a finite radius of convergence. The Borel summation of $f^{(0)}(w)$ is then formally defined as  
\bear
{\cal S}_\theta f^{(0)}(w)=f^{(0)}_0+\int_0^{e^{i \theta}\infty} e^{-ws}\,{\hat f^{(0)}}(s)\,ds\,.
\label{eq:borel-resum}
\eear
The angle $\theta$ determines the contour of integration in the complex- $s$ plane, usually called the ``Borel plane", and is correlated with the phase of the original expansion parameter $w$.
For our original problem we are interested in $\theta=0$, since $w$ is a real parameter. However 
${\hat f^{(0)}(s)}$ has singularities along the real axis, where $\theta=0$. A quick way to see this is to use the large order expression \eqref{eq:zero-growth} in \eqref{eq:borel-def}:
\begin{equation}
{\hat f^{(0)}}(s)\sim{S_1\over 2\pi i} \,\Gamma(\beta+1)\left(S-s\right)^{-\beta-1} \left[f^{(1)}_0+{S-s\over \beta}f^{(1)}_0+{(S-s)^2\over \beta(\beta-1)}f^{(1)}_0+\dots\right]+\dots\,.
\label{eq:quick-borel}
\end{equation}
It is clear that there is a branch point at $s=S$. In order to avoid the branch cut we deform the contour infinitesimally as $\theta=0^{\pm}$, but depending on the direction, one encounters a discontinuity.  Let us consider $\theta=0^+$. The discontinuity is pure imaginary, and is of the order ${\cal I}m [f^{(0)}(w)] \propto S_1 f^{(1)}_0 \,w^\beta e^{-S\,w}$. Requiring that the trans-series is real, this ambiguous contribution to $f^{(0)}(w)$ that arises from the singularity in the Borel transform must be canceled by $f^{(1)}(w)$. This cancellation fixes the imaginary part of $\sigma$ in the trans-series \eqref{eq:trans2} to be \cite{Marino:2008ya,Aniceto:2013fka}
\begin{eqnarray}
 {\cal I}m[\sigma]=-\frac{1}{2} \, {\cal I}m[S_1] 
 \label{eq:canc}
 \end{eqnarray}
With ${\cal N}=4$ parameters, ${\cal I}m(S_1) \approx -0.04$. For a detailed analysis of the reality of trans-series, and the associated cancellations of imaginary ambiguities we refer the reader to \cite{Aniceto:2013fka}.

\begin{figure}[h]
\includegraphics[width=0.65\textwidth]{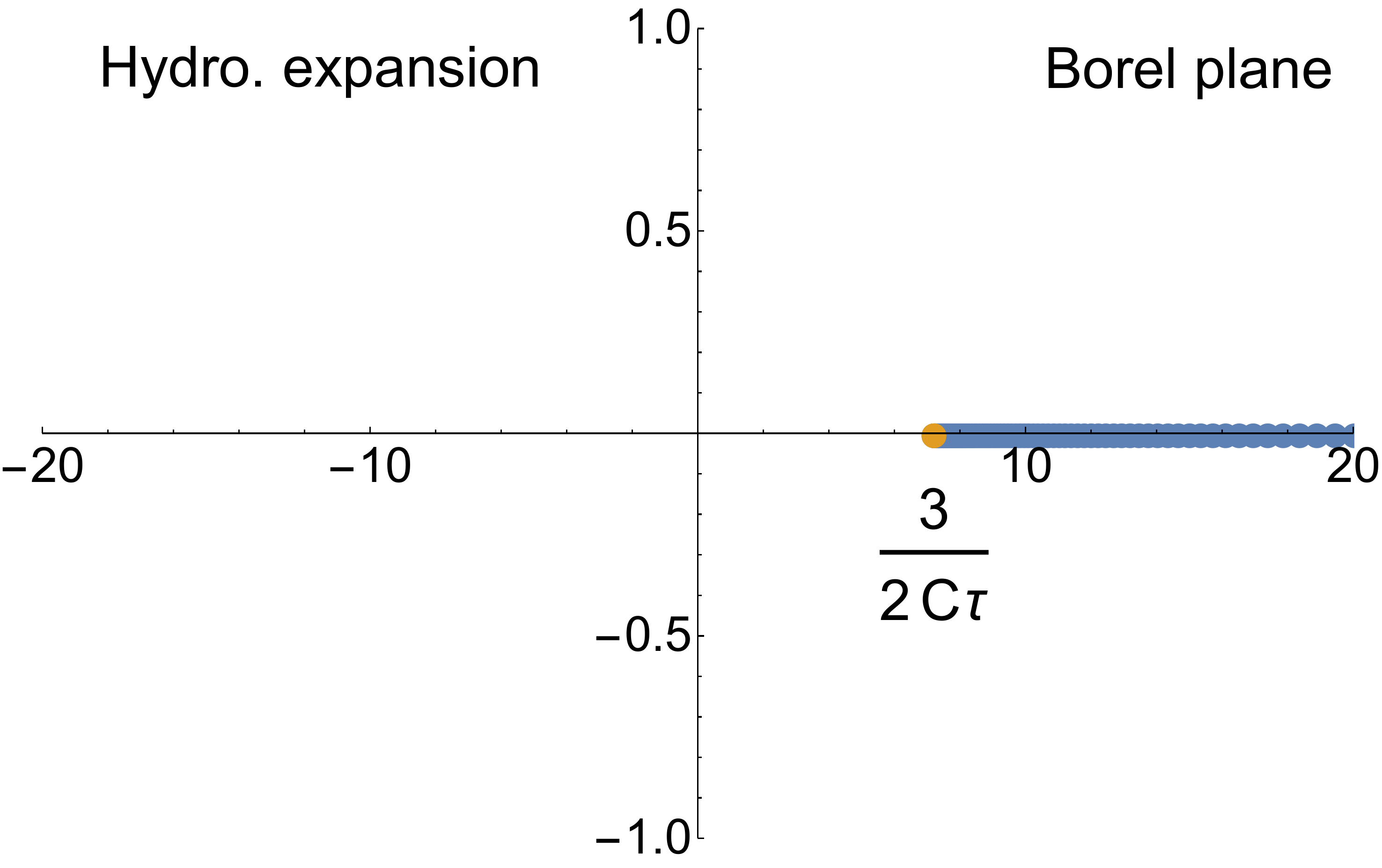}
\caption{The poles in the Borel complex $s$-plane of the Pad\'e approximant to the Borel transform of the hydrodynamic expansion, ${\hat f^{(0)}}(s)$. The poles accumulate into a branch cut that starts at $S={3\over2C_\tau}$, which is the analog of the instanton action.  }
\label{fig:pade-hydro}
\end{figure}

The branch cut singularity of the Borel transform ${\hat f^{(0)}}(s)$ can also be seen in another way. By generating many coefficients $f^{(0)}_k$, using the recursion relations \eqref{eq:rec-pert},  we automatically generate many terms in the Taylor series for ${\hat f^{(0)}}(s)$, from (\ref{eq:borel-def}). The location of the nearest singularity may be deduced by a root test or ratio test, but to see the branch cut it is better to use a Pad\'e approximant \cite{bender}. The Pad\'e approximant to ${\hat f^{(0)}}(s)$ approximates it as a ratio of two polynomials:
\bear
{\hat f^{(0)}}(s)\approx {p^{(0)}(s)\over q^{(0)}(s)}\,.
\eear
The branch cuts of the actual Borel transform function manifest themselves as an accumulation of poles in the Pad\'e approximant. Of course the further away a branch point is from the origin, the more precision and number of terms it would require to reproduce the associated branch cut. We computed the symmetric Pad\'e approximant of order 300; namely $p^{(0)}(s)$ and $q^{(0)}(s)$ are taken to be polynomials of order 150.  We used precision of 800 significant figures. The result is shown in Fig. \ref{fig:pade-hydro}. It is clear that there is a branch cut that starts at $s=S={3\over 2 C_\tau}$, along the positive real axis, consistent with the resurgent behavior in (\ref{eq:quick-borel}) and the large-order behavior in (\ref{eq:zero-growth}). 
The fact that the location of the singularity in the Borel plane coincides with the ``instanton action'' \eqref{eq:xi}  in the trans series is a generic feature of resurgence. 

Due to the nonlinear nature of the original differential equation (\ref{eq:main-eqn}),  there are in fact infinitely many exponentially suppressed terms in the trans-series, each of which is associated with an action $n S$. This fact translates into the existence of branch points located at $s=n S$ for all $n\geq1$. The ambiguities that arise from each of these branch points are cured by the existence of the $n^{th}$ non-hydrodynamic series. In terms of the large order growth this means that the late terms of the hydrodynamic series actually contains information about \textit{all} the non-hydrodynamic series. More precisely, the expression \eqref{eq:zero-growth} can be further refined as  \cite{Aniceto:2013fka}
\bear
f^{(0)}_k&\sim&\, S_1\,{\Gamma(k+\beta)\over 2 \pi i \,S^{k+\beta}}\left(f^{(1)}_0+{S\over k+\beta-1}\,f^{(1)}_1+{S^2\over (k+\beta-1)(k+\beta-2)}\,f^{(1)}_2+\dots\right)\nn
&&+S_1^2\,{\Gamma(k+2\beta)\over 2 \pi i \,(2S)^{k+\beta}}\left(f^{(2)}_0+{2S\over k+2\beta-1}\,f^{(2)}_1+{(2S)^2\over (k+2\beta-1)(k+2\beta-2)}\,f^{(2)}_2+\dots\right)\nn
&&+S_1^3\,{\Gamma(k+3\beta)\over 2 \pi i \,(3S)^{k+\beta}}\left(f^{(3)}_0+{3S\over k+3\beta-1}\,f^{(3)}_1+{(3S)^2\over (k+3\beta-1)(k+3\beta-2)}\,f^{(3)}_2+\dots\right)\nn
&&+\dots\,.
\label{eq:zero-growth2}
\eear
Note that the expression in each line involves low order coefficients of non-hydrodynamic series of different order, and that there is only one constant,  the Stokes constant $S_1$, that needs to be determined numerically.

\subsection{First (leading) non-hydrodynamic expansion: $f^{(1)}(w)$}
\label{sec:first_expansion}
\begin{figure}[h]
\includegraphics[width=0.495\textwidth]{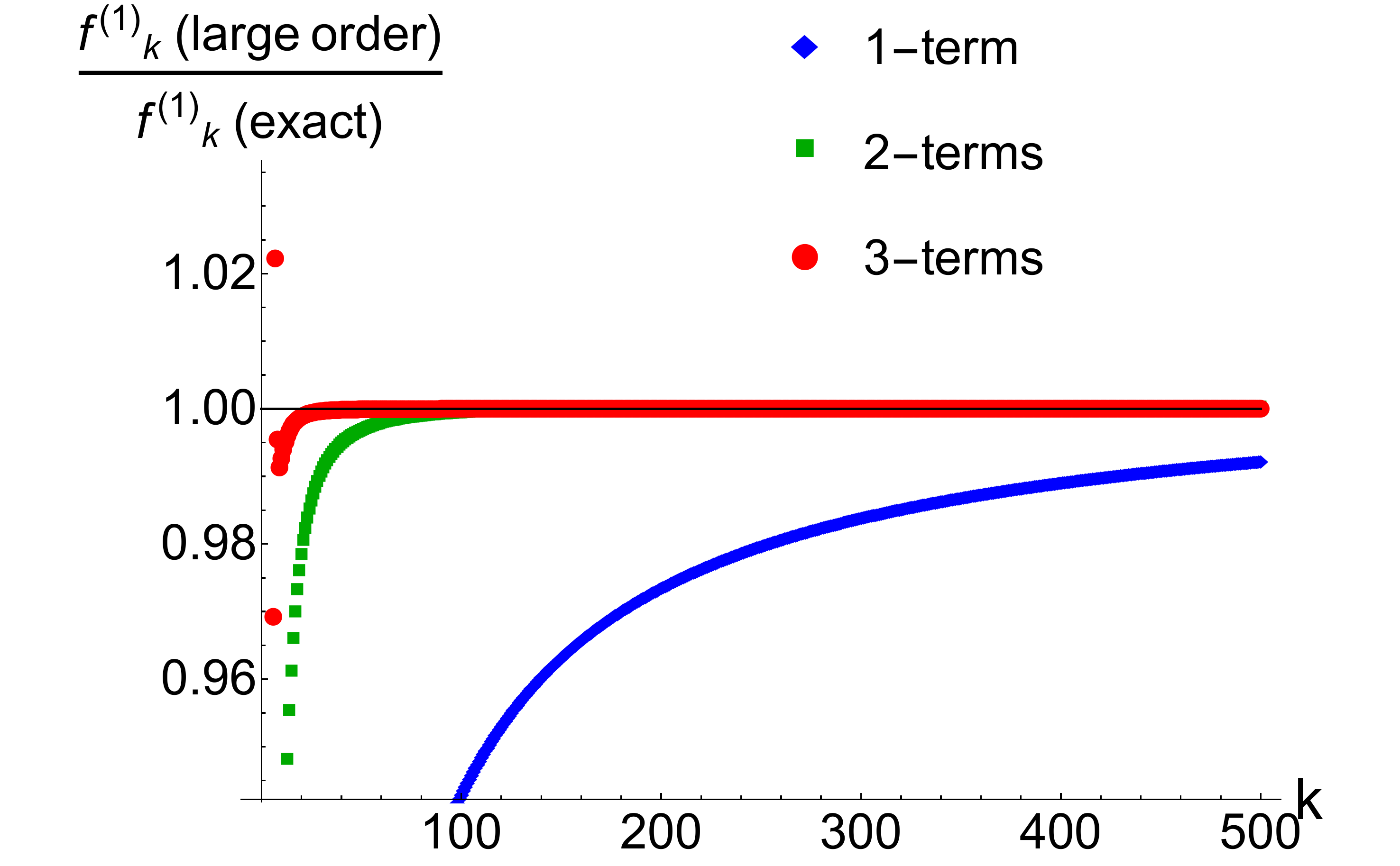}
\includegraphics[width=0.495\textwidth]{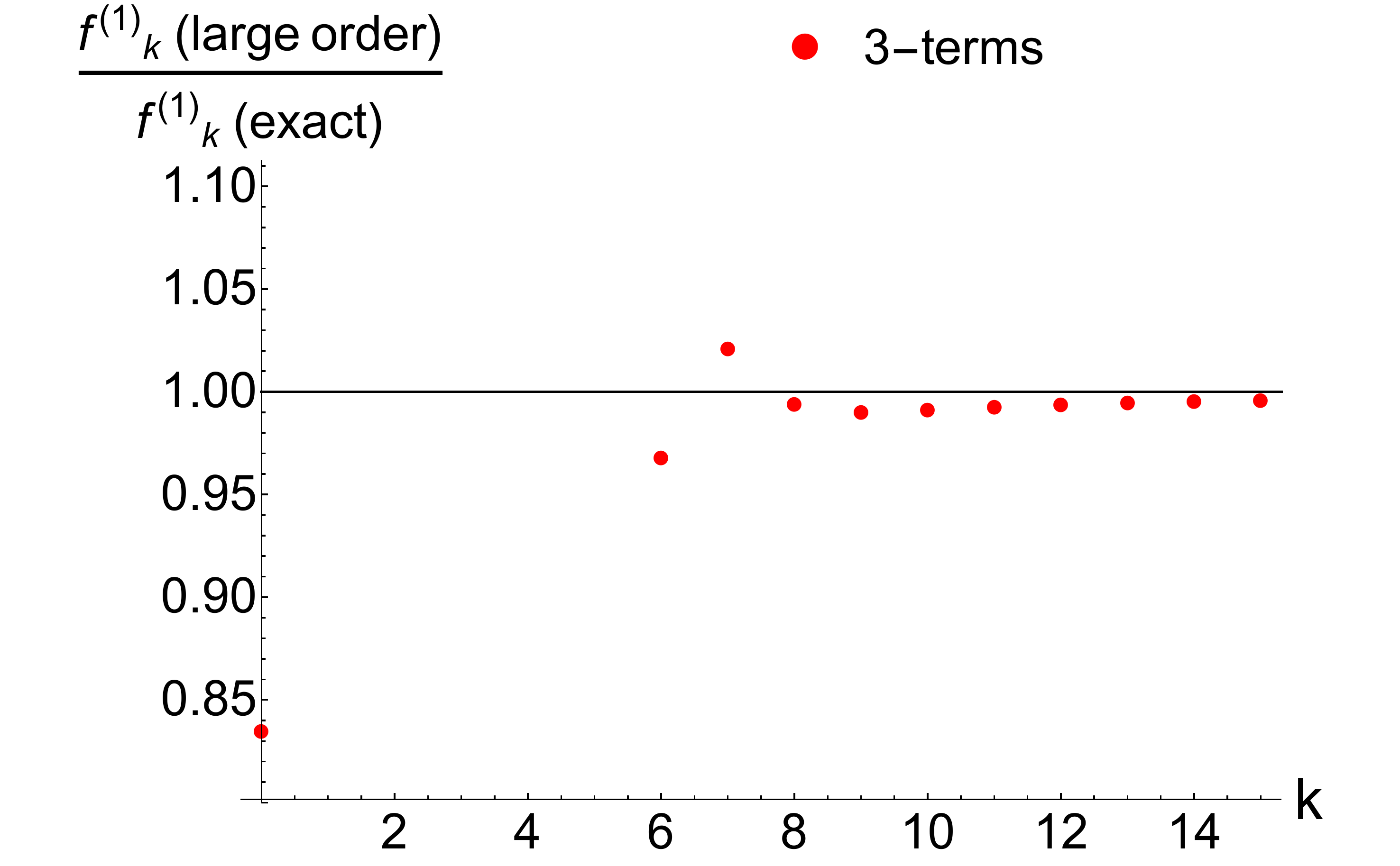}
\caption{The first plot shows plot the ratio of the large order expression \eqref{eq:first-growth} to the exact coefficients $f^{(1)}_k$, as a function of $k$. The blue, green and red points refer to the inclusion of the zeroth, first, and second subleading terms in \eqref{eq:first-growth}. The second plot shows a close-up view of the ratio using the  expression with the first three terms in \eqref{eq:first-growth}. This clearly demonstrates the remarkable precision of the resurgence relation \eqref{eq:first-growth}:  the agreement is at the one percent level already by $k\approx 10$.}
\label{fig:1-lo}
\end{figure}

We can repeat the analysis of the previous section for the fluctuations $f^{(1)}(w)$ about the first non-hydrodynamic term in the trans-series. Once again, general arguments for a real trans-series predict the large order growth  \cite{Aniceto:2013fka}:
\bear
f^{(1)}_k\sim 2 S_1\,{\Gamma(k+\beta)\over 2 \pi i \,S^{k+\beta}}\left(f^{(2)}_0+{S\over k+\beta-1}\,f^{(2)}_1+{S^2\over (k+\beta-1)(k+\beta-2)}\,f^{(2)}_2+\dots\right)+\dots
\label{eq:first-growth}
\eear
The coefficients $f^{(2)}_k$ on the right-hand side are the low order terms of the second non-hydrodynamic series $f^{(2)}(w)$ in (\ref{eq:f2}). Note that in (\ref{eq:first-growth}) all constants on the right-hand-side are known:  the overall normalization constant is fixed to be twice the very same Stokes constant $S_1$ found in the large-order behavior \eqref{eq:zero-growth} of the hydrodynamic series. In this  sense \eqref{eq:first-growth} is an even stronger prediction than \eqref{eq:zero-growth}. In Figure \ref{fig:1-lo} we plot the ratio of the large order expression \eqref{eq:first-growth} to the exact coefficients generated from the recursion relations (\ref{eq:rec-1}). Again, the agreement is excellent.
 
\begin{figure}[h]
\includegraphics[width=0.65\textwidth]{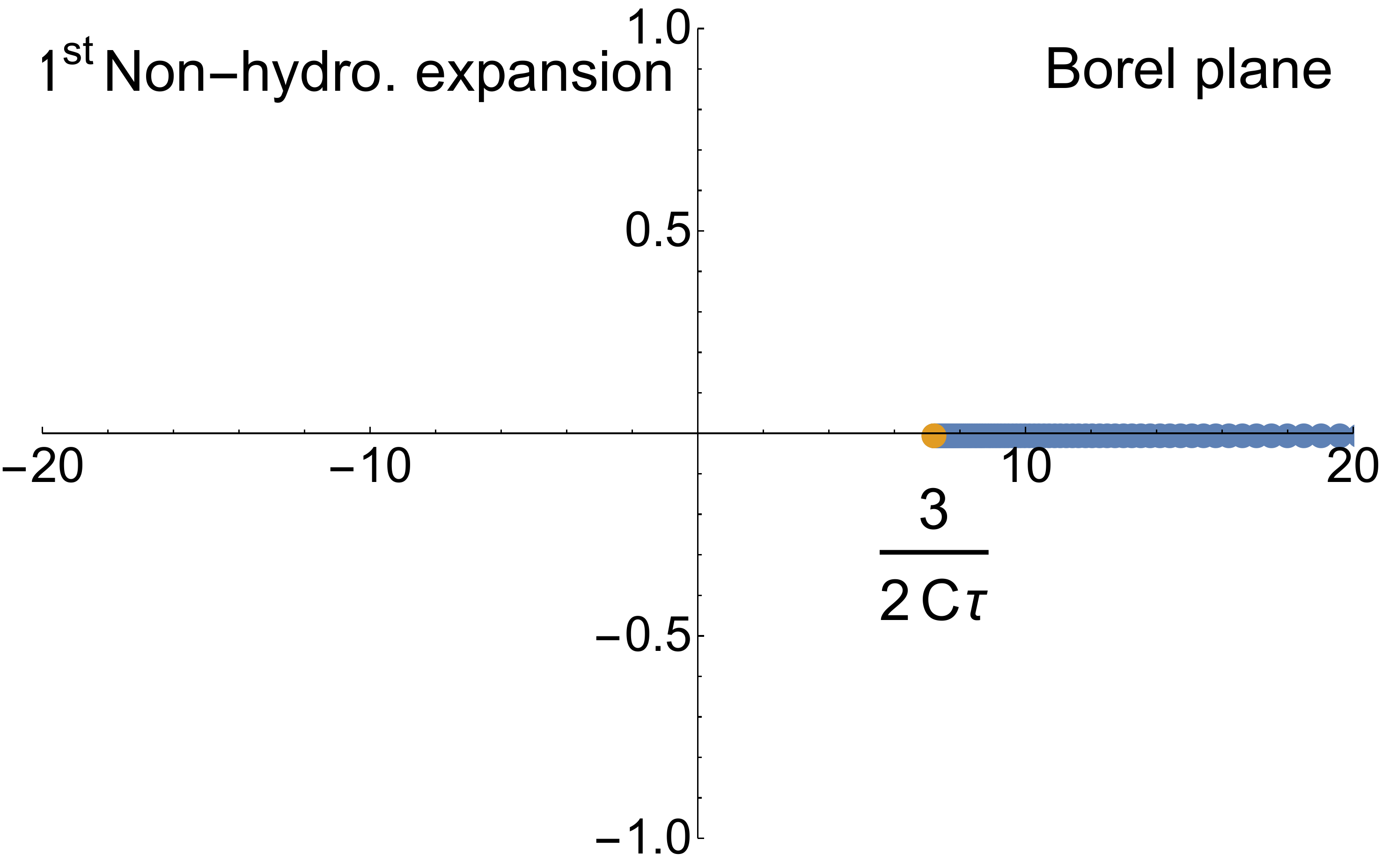}
\caption{The poles in the Borel complex $s$-plane of the Pad\'e approximant to the Borel transform of the first non- hydrodynamic expansion, ${\hat f^{(1)}}(s)$. The poles accumulate into a branch cut that starts at $S={3\over2C_\tau}$.}
\label{fig:pade-1}
\end{figure}

 The Borel plane structure can be deduced by the same Borel-Pad\'e analysis, and as shown in Figure \ref{fig:pade-1} we see once again a cut along the positive real axis, starting at the branch point $S$.
The Borel plane structure is almost identical to that of the hydrodynamic series; namely there are infinitely many branch cuts located at $s=nS$ and accordingly, the large order growth \eqref{eq:first-growth} can be refined similar to \eqref{eq:zero-growth2}.  In other words, late terms of $f^{(1)}(w)$ contain the low order terms of all the higher non-hydrodynamic series $f^{(n)}(w)$ with $n\geq2$. 

\subsection{Second (next-to-leading) non-hydrodynamic expansion: $f^{(2)}(w)$}
\label{sec:second_expansion}

At the next order, studying the large order growth of the coefficients of the second non-hydrodynamic series $f^{(2)}(w)$, we observe a new phenomenon, the possibility of which was pointed out by Aniceto and Schiappa in their exhaustive analysis of the resurgent structure of real trans-series \cite{Aniceto:2013fka}. The large order growth of the 
$f^{(2)}_k$ coefficients is determined not only by the the low order coefficents of $f^{(3)}(w)$, but also by the the low order coefficents of $f^{(1)}(w)$:
\bear
f^{(2)}_k&\sim&S_{-1}\,{\Gamma(k-\beta)\over 2 \pi i \,(-S)^{k-\beta}}\left(f^{(1)}_0+{-S\over k-\beta-1}\,f^{(1)}_1+{(-S)^2\over (k-\beta-1)(k-\beta-2)}\,f^{(1)}_2+\dots\right)+\nn
&&3 S_1\,{\Gamma(k+\beta)\over 2 \pi i \,S^{k+\beta}}\left(f^{(3)}_0+{S\over k+\beta-1}\,f^{(3)}_1+{S^2\over (k+\beta-1)(k+\beta-2)}\,f^{(3)}_2+\dots\right)\nn
&&+\dots\,. 
\label{eq:second-growth}
\eear
Here $S_{-1}$ is a new Stokes constant whose value we determine numerically to be $S_{-1}\approx  -57.922+115.651 i$. Note that in the first line the phase of $S_{-1}$ is cancelled by the factor $ i (-1)^{-\beta}$ in the denominator, as  the coefficients $f^{(2)}_k$ are real. 

\begin{figure}[h]
\includegraphics[width=0.495\textwidth]{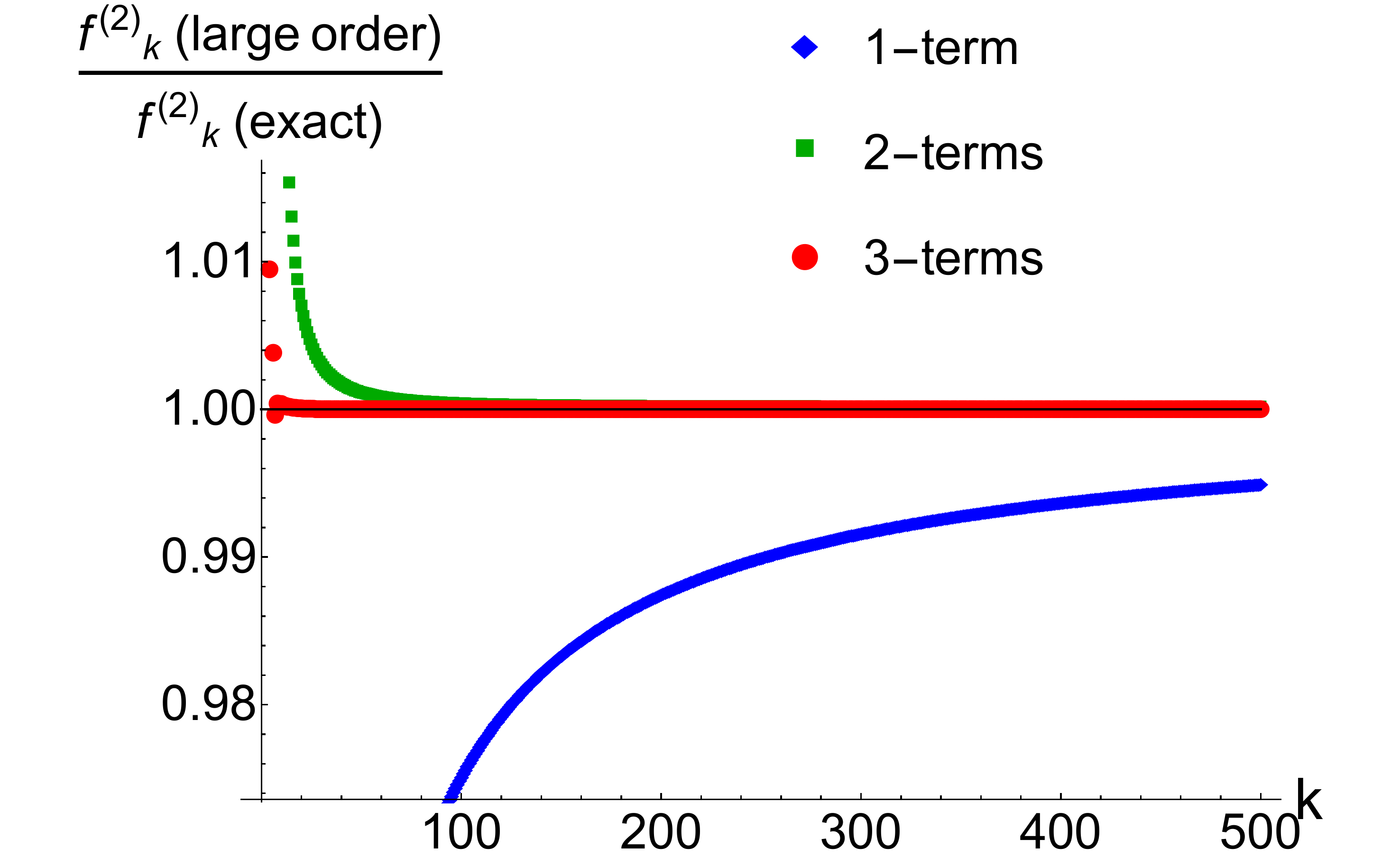}
\includegraphics[width=0.495\textwidth]{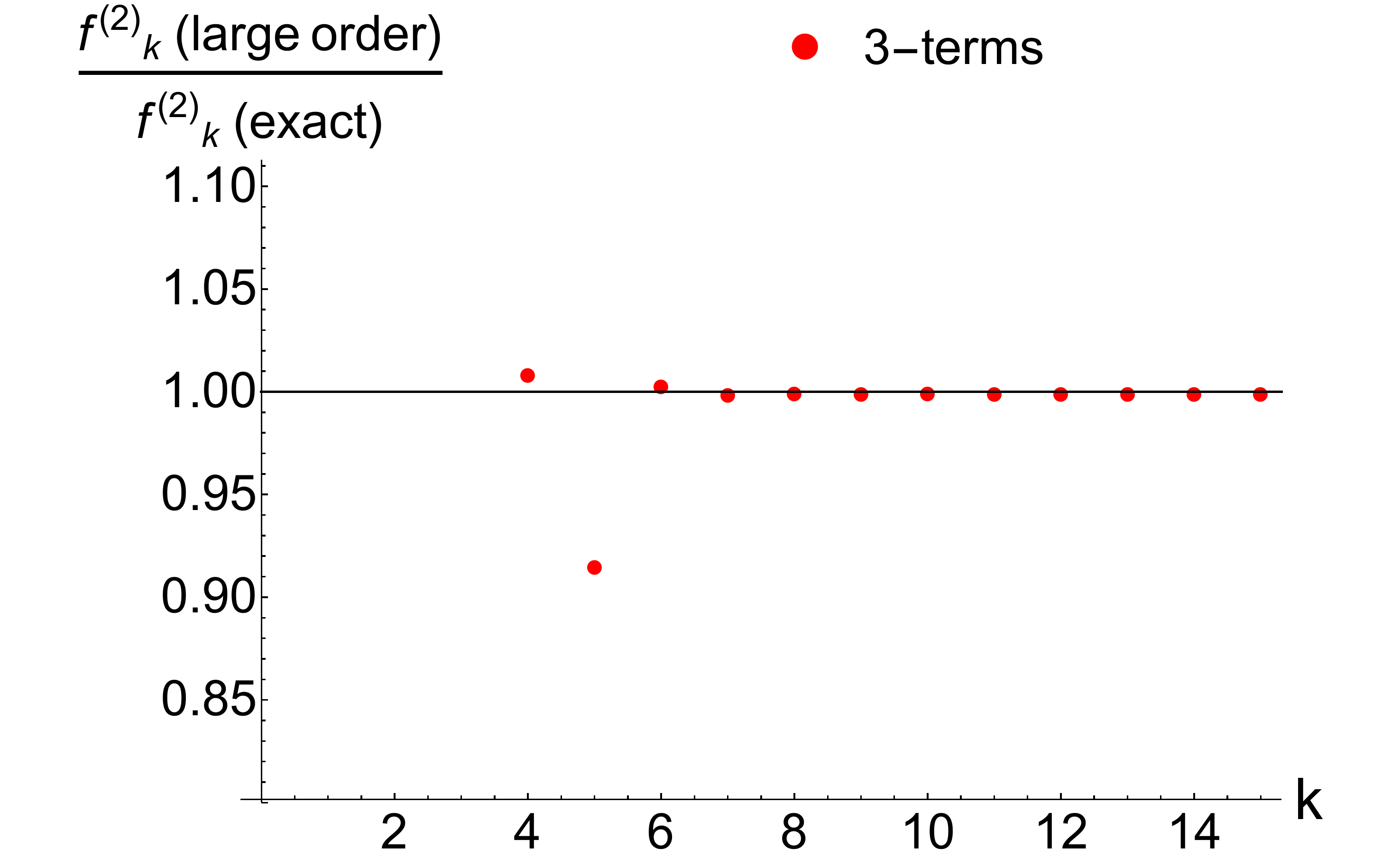}
\caption{The first plot shows plot the ratio of the large order expression \eqref{eq:second-growth} to the exact coefficients $f^{(2)}_k$, as a function of $k$. The blue, green and red points refer to the inclusion of the zeroth, first, and second subleading terms in \eqref{eq:second-growth}. The second plot shows a close-up view of the ratio using the  expression with the first three terms in \eqref{eq:second-growth}. This clearly demonstrates the remarkable precision of the resurgence relation \eqref{eq:second-growth}:  the agreement is at the one percent level already by $k\approx 10$.}
\label{fig:2-lo}
\end{figure}

\begin{figure}[h]
\includegraphics[width=0.65\textwidth]{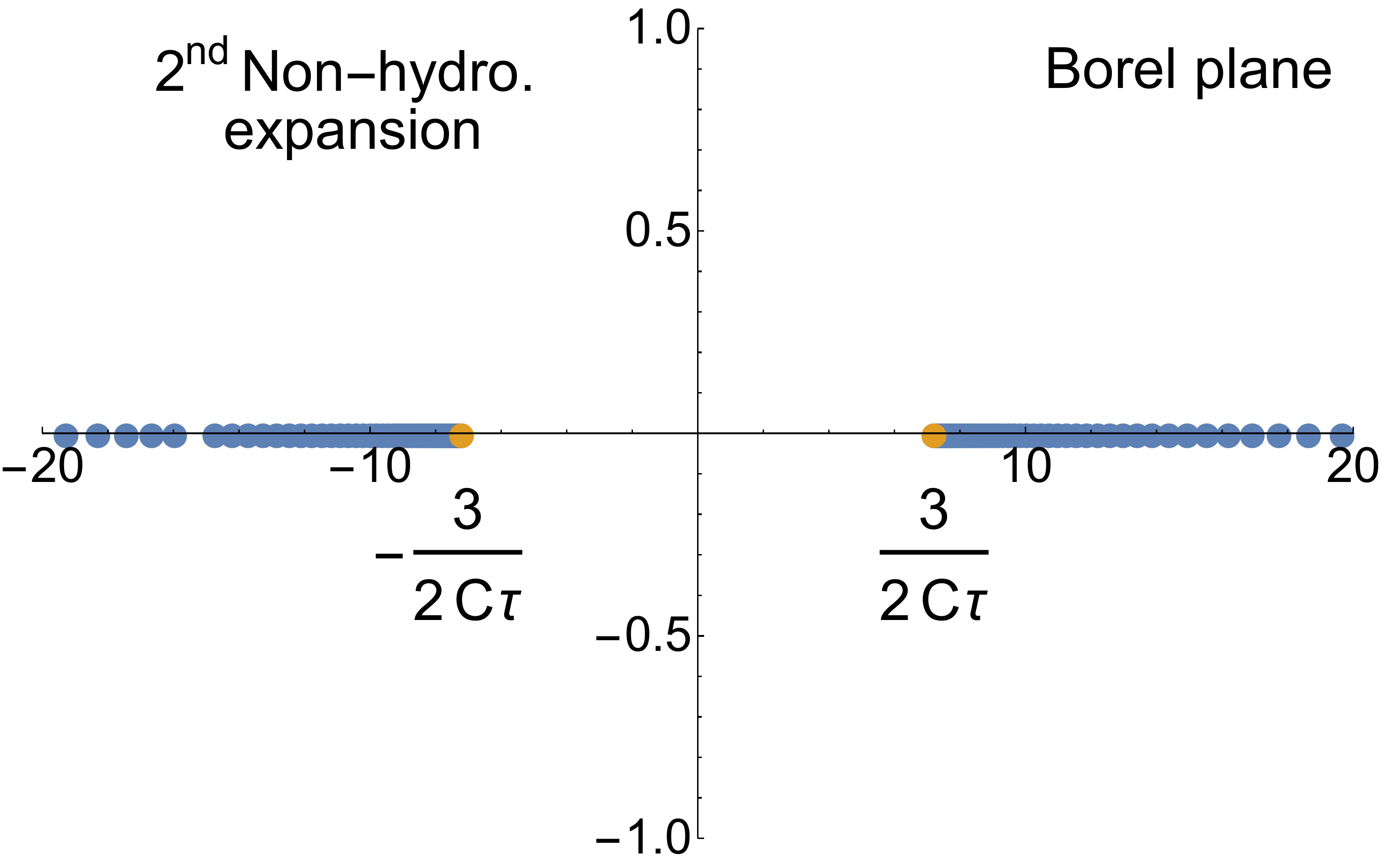}
\caption{The poles in the Borel complex $s$-plane of the Pad\'e approximant to the Borel transform of the second non-hydrodynamic expansion, ${\hat f^{(2)}}(s)$. The poles accumulate into two branch cuts, which start at $S={3\over2C_\tau}$, and at $S=-{3\over2C_\tau}$.  }
\label{fig:pade-2}
\end{figure}

In Fig \ref{fig:2-lo} we plot the ratio of the expression \eqref{eq:second-growth} to the exact coefficients $f^{(2)}_k$ generated from the recursion relations in (\ref{eq:rec-n}). As in the previous cases, the agreement is excellent. Note that if we do not include the effect of the $f^{(1)}_k$ coefficients, the agreement is terrible. In fact, for the particular choice of ${\cal N}=4$ parameters that we study, because $\beta<0$, the contribution of the first non-hydrodynamic sector (i.e. the first line in \eqref{eq:second-growth}) is more dominant compared to the third non-hydrodynamical sector (the second line in \eqref{eq:second-growth}). Related with this fact, the coefficients of the series $f^{(2)}(w)$ are actually sign-alternating, as opposed to the hydrodynamic series $f^{(0)}(w)$ and the first non-hydrodynamic series $f^{(1)}(w)$, for which the expansion coefficients are non-alternating.  
But this sign-alternating behavior  does not necessarily mean that $f^{(2)}(w)$ is Borel summable. To explore this, we use the Borel-Pad\'e method to study the singularity structure of the Borel transform $\hat{f}^{(2)}(s)$.
In the complex-$s$ Borel plane (see Fig. \ref{fig:pade-2}), in addition to the infinitely many branch points along the positive real axis $s=nS$, associated with the higher ($3^{rd}$, $4^{th}$, $5^{th}$, etc.) non-hydrodynamic series, there is an additional branch cut along the \textit{negative} real axis that starts at the branch point $s=-S$. This branch cut is associated with the first non-hydrodynamic series, and the corresponding Stokes constant $S_{-1}$  is related to the Stokes discontinuity at $\theta=\pi$.  
It is clear from the existence of the branch cuts on the positive real axis, $f^{(2)}(w)$ is not Borel summable along the positive real axis, even though it is an alternating series. A better way to conceptualize this is to view $f^{(2)}(w)$ as being composed of different sub-series, each of which is related to different non-hydrodynamic sectors (each line in \eqref{eq:second-growth}). The sub-series related to the first non-hydrodynamic series is alternating, Borel summable along the positive real axis, and is the dominant contribution to the late term coefficients $f^{(2)}_k$. Therefore the whole series $f^{(2)}(w)$ is alternating. However, all the sub-series associated with non-hydrodynamic series with $n\geq3$ lead to non-alternating sub-series which are not Borel summable along the positive real axis.  This phenomenon is reminiscent of the quantum mechanical example discussed in \cite{Basar:2013eka},  where there are complex instantons with negative actions, dubbed  ``ghost instantons'', which contribute to the large-order behavior of perturbation theory; the structure is very similar to the way in which the first non-hydrodynamic series contributes to the large order growth of the $f^{(2)}_k$ expansion coefficients.

\subsection{Third and higher non-hydrodynamic expansions: $f^{(n)}(w)$ with $n\geq 3$}
\label{sec:third_expansion}

\begin{figure}[h]
\includegraphics[width=0.495\textwidth]{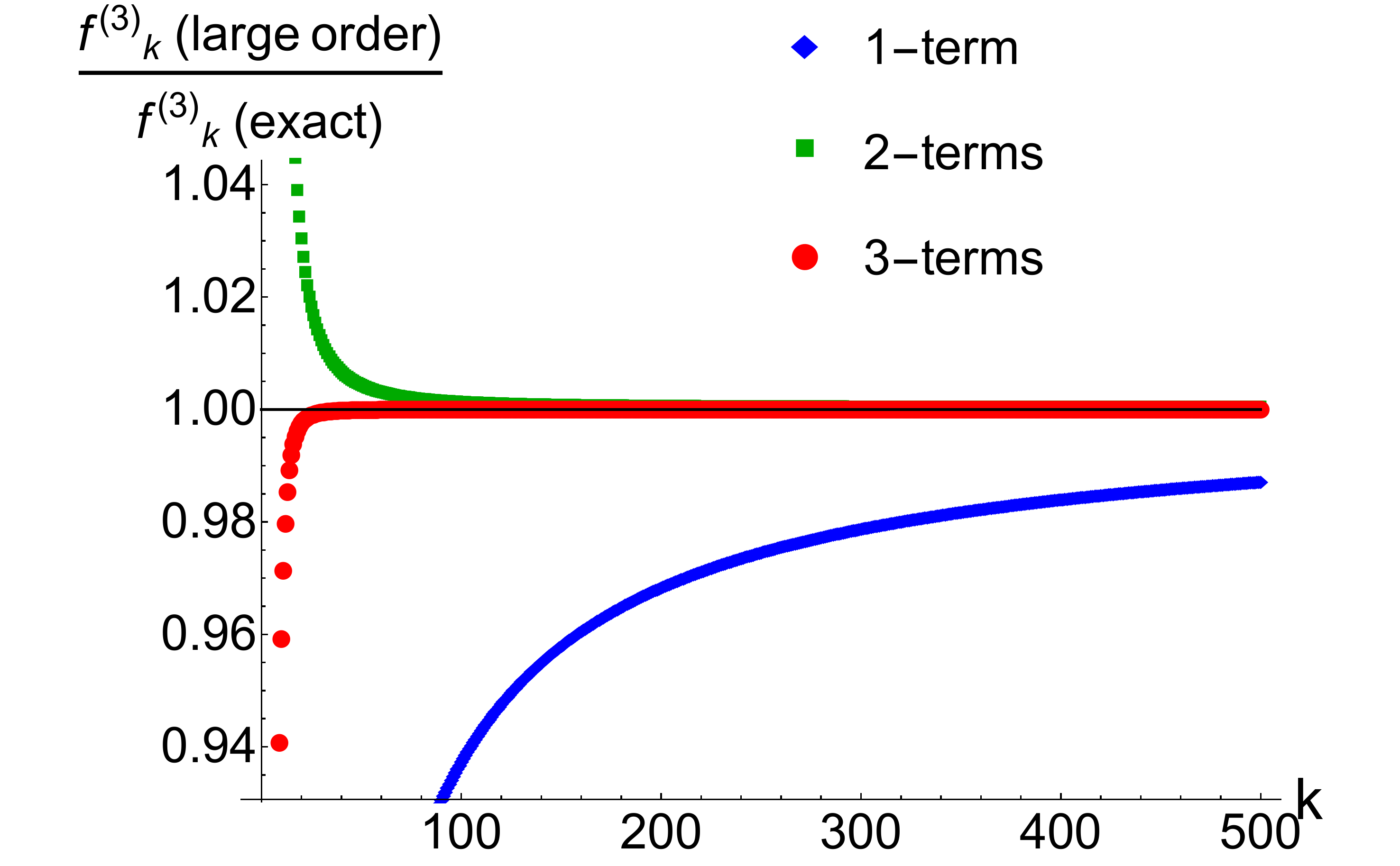}
\includegraphics[width=0.495\textwidth]{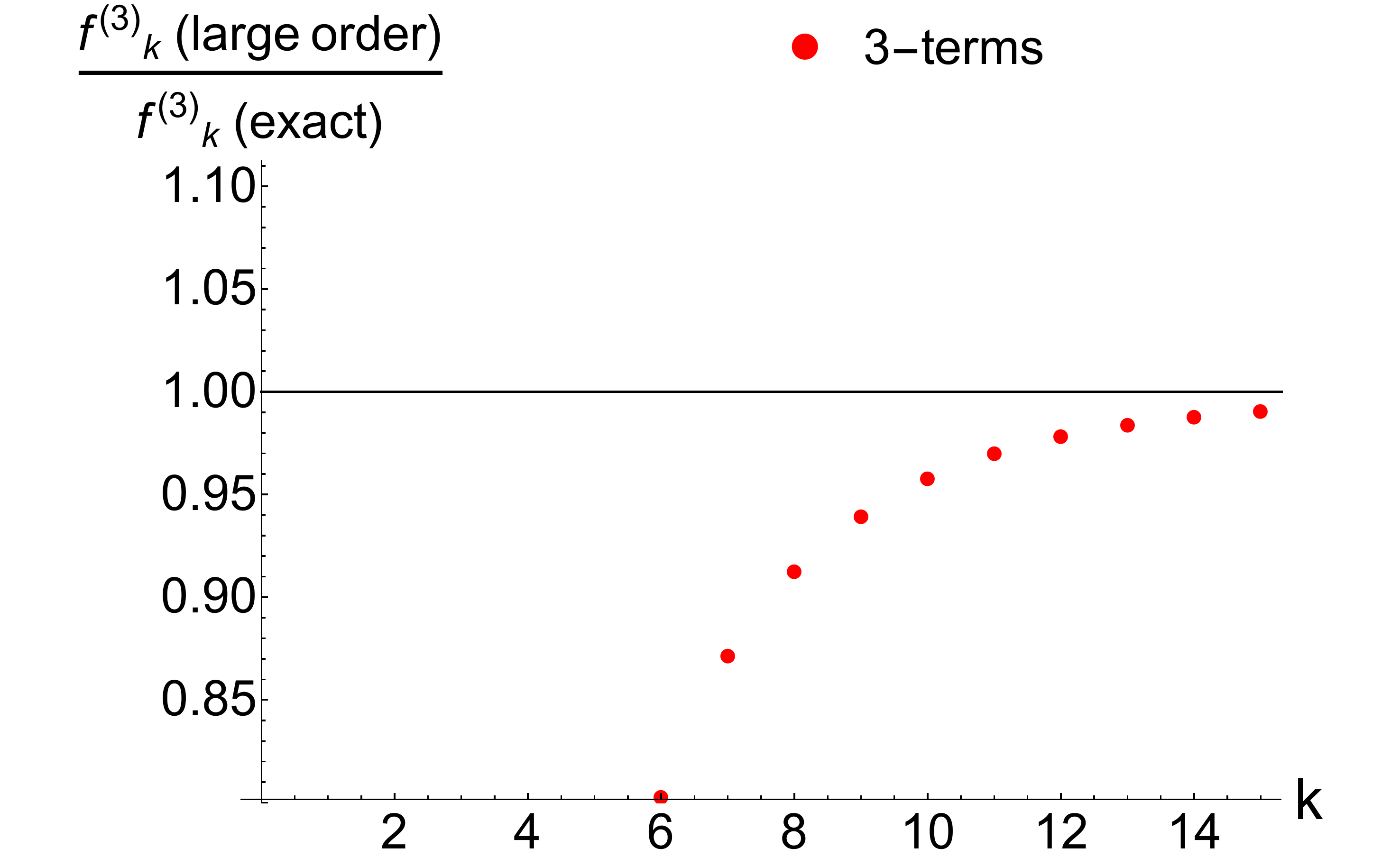}
\caption{The first plot shows plot the ratio of the large order expression \eqref{eq:third-growth} to the exact coefficients $f^{(3)}_k$, as a function of $k$. The blue, green and red points refer to the inclusion of the zeroth, first, and second subleading terms in \eqref{eq:third-growth}. The second plot shows a close-up view of the ratio using the  expression with the first three terms in \eqref{eq:third-growth}. This clearly demonstrates the remarkable precision of the resurgence relation \eqref{eq:third-growth}:  the agreement is at the few-percent level already by $k\approx 10$.}
\label{fig:3-lo}
\end{figure}

\begin{figure}[h]
\includegraphics[width=0.65\textwidth]{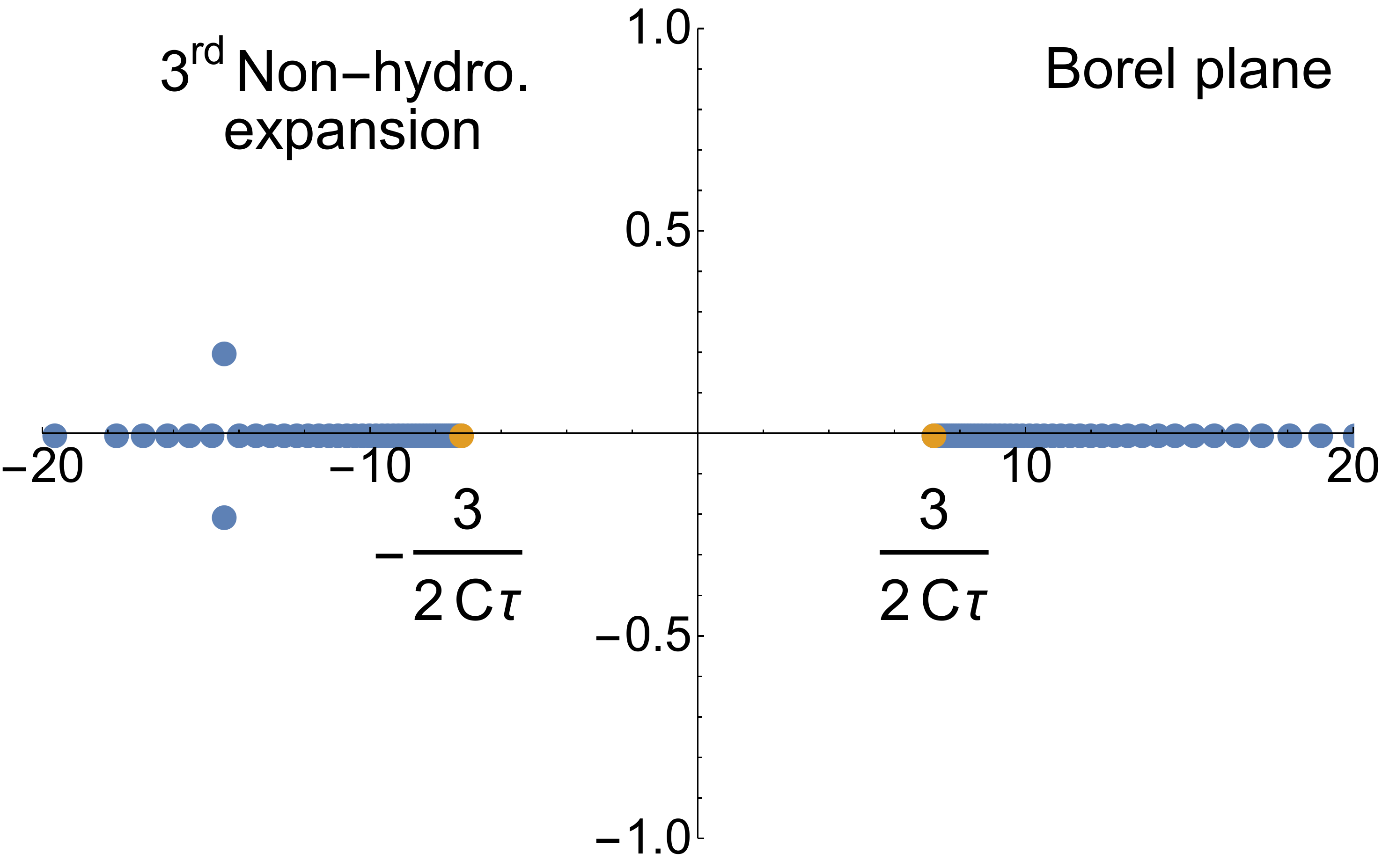}
\caption{The poles in the Borel complex-$s$ plane of the Pad\'e approximant to the Borel transform of the third non-hydrodynamic expansion, ${\hat f^{(3)}}(s)$. The poles accumulate into a two branch cuts that start at $s={3\over 2C_\tau}$ and $s=-{3\over 2C_\tau}$. The two poles that do not lie on the real  axis are artifacts of the finite order (150) of our Pad\'e approximation.}
\label{fig:pade-3}
\end{figure}

The large-order behavior found in the second non-hydrodynamic sector in the previous section persists at third order and beyond. For example, at third order we find the large-order growht of the expansion coefficients:
\bear
f^{(3)}_k&\sim&2S_{-1}\,{\Gamma(k-\beta)\over 2 \pi i \,(-S)^{k-\beta}}\left(f^{(2)}_0+{-S\over k-\beta-1}\,f^{(2)}_1+{(-S)^2\over (k-\beta-1)(k-\beta-2)}\,f^{(2)}_2+\dots\right)+\dots\nn
&&4 S_1\,{\Gamma(k+\beta)\over 2 \pi i \,S^{k+\beta}}\left(f^{(4)}_0+{S\over k+\beta-1}\,f^{(4)}_1+{S^2\over (k+\beta-1)(k+\beta-2)}\,f^{(4)}_2+\dots\right)+\dots\nn
\label{eq:third-growth}
\eear
with the same $S_{-1}$ parameter. The agreement of (\ref{eq:third-growth}) with the exact coefficients generated from the recursion relations (\ref{eq:rec-n}) is again excellent, as shown in Figure \ref{fig:3-lo}. We stress that there are no free parameters in this comparison. All constants on the right-hand-side of (\ref{eq:third-growth}) are determined.
The Borel plane structure of the Borel transform ${\hat f^{(3)}}(s)$ is shown in Figure \ref{fig:pade-3}, showing the same two branches, starting at $s=\pm {3\over 2C_\tau}$.
We expect that this Borel plane structure persists to all orders.

For completeness, we note that there are further even smaller corrections to (\ref{eq:third-growth}), involving $f^{(1)}$, and also $f^{(n)}$ with $n\geq 5$. The $f^{(1)}$ part introduces a new Stokes constant $S_{-2}$ which can be determined numerically. The general pattern for the large order growth of the $N^{th}$ hydrodynamical series $f^{(N)}(w)$ is as follows. There is a set of alternating sub-series related to the sectors $n=1,2,\dots,N-1$, each of which has an associated Stokes constant $S_{-(N-n)}$. They generate branch points in the Borel plane along the negative real axis: $s=-S,-2S,\dots,-(N-1)S$. As mentioned before, there is another set of sub-series related to the sectors $n=N+1,N+2,\dots$ all of which are non-alternating. They generate branch cuts at $s=S,2S,\dots$. In other words the location of the branch cut in the Borel plane of ${\hat f^{(N)}}(s)$ due to the non-hydrodynamical sector $n\neq N$ is given by the ``relative action" $s_{branch}=(n-N) S$. This is an illustration of the general behavior of real trans-series and has also been observed in resurgent analysis of topological string theory   \cite{Pasquetti:2009jg,Aniceto:2013fka}.

\section{Trans-asymptotic rearrangement and initial conditions}
\label{sec:trans_asympt}

At earlier proper-time (i.e. smaller $w$), the terms in the trans-series expansion (\ref{eq:trans-ansatz}) become disordered, as exponentials compete in size with inverse  powers of $w$. In such a situation, there is a systematic way to rearrange the terms of the  trans-series as one approaches the edge of its domain of numerical usefulness \cite{costin,costin-costin}. This is the first step in ``trans-asymptotic matching'' \cite{costin-costin}.  This procedure effectively  resums all the exponentially small non-hydrodynamic modes for a given power of $1/w$, and rearranges the trans-series (\ref{eq:trans2})  into the form 
\bear
f(w)\sim\sum_{k=0}^\infty {F_k(\sigma\, \zeta) \over w^k}
\label{eq:transasymptotic}
\eear
This means that we have identified $F_k(\sigma\, \zeta)$ with the formal expansion 
\bear
F_k(\sigma\, \zeta) = \sum_{n=0}^\infty f_k^{(n)}(w) \, \sigma^n\, \zeta^n
\label{eq:transF}
\eear
but we will see that we can in fact obtain closed-form expressions for the $F_k(\sigma\, \zeta)$.

In (\ref{eq:transasymptotic}),  $\zeta=\zeta(w)$ is the same  ``instanton fugacity'', $\zeta(w)\equiv w^\beta \,e^{-Sw}$, introduced in (\ref{eq:xi}). Inserting  this reorganized ansatz for $f(w)$  into the original differential equation (\ref{eq:main-eqn}), leads to a sequence of ordinary differential equations, one for each $F_k(\zeta)$, with $\zeta$ now being regarded as the independent variable. Note that the argument of $F_k$ involves also an arbitrary numerical factor multiplying $\zeta$. As in the expansion  (\ref{eq:trans2}), there is just one such undetermined constant, and by comparison we identify it with the trans-series parameter $\sigma$ in (\ref{eq:trans2}).  The tower of differential equations for $F_k(\zeta)$ can be solved recursively: the first equation of the tower involves only $F_0(\zeta)$, the second only involves $F_0(\zeta)$ and $F_1(\zeta)$, and so on. [For notational convenience we scale $\sigma=1$ here, and reintroduce it again later]. Moreover, while the equation for $F_0(\zeta)$ is nonlinear, all subsequent equations for $F_k(\zeta)$ are linear. Furthermore, when expanded as series in $\zeta$, each function $F_k(\zeta)$ is {\it convergent}, even though the full trans-series expression (\ref{eq:transasymptotic}) is of course still divergent \cite{costin-costin}.

 It is straightforward to show that the ansatz (\ref{eq:transasymptotic}) leads to the following equations for $F_k(\zeta)$:
\bear
&&C_\tau\sum_{k^\prime=0}^k\left(-S\zeta {d F_{k^\prime} \over d\zeta} +\beta \zeta {dF_{k^\prime-1}\over d\zeta}-(k^\prime-1)F_{k^\prime-1}(\zeta)\right) F_{k-k^\prime}(\zeta)+{3\,C_\lambda\over 2\,C_\eta}\sum_{k^\prime=0}^kF_{k^\prime}(\zeta)F_{k-k^\prime}(\zeta)
\nn
&&+C_\tau\left(-S\zeta {d F_{k} \over d\zeta} +\beta \zeta {dF_{k-1}\over d\zeta}-(k-1)F_{k-1}(\zeta)\right) +{8\over3} C_\tau F_{k-1}(\zeta)+4 C_\tau \sum_{k^\prime=0}^{k-1} F_{k-k^\prime-1}(\zeta)F_{k^\prime}(\zeta)
\nn
&&
+\left(1+{C_\lambda\over C_\eta}\right)F_k(\zeta)+{4\over 9}( C_\tau- C_\eta) \delta_{k-1,0}+{2\over 3}\left( 1+{C_\lambda \over2\, C_\eta}\right)\delta_{k,0}=0
\label{eq:tower}
\eear 
We focus on the class of parameters with $C_\lambda=2C_\eta$, which includes the particular case of  $\mc N=4$ parameters that we have been using for numerical purposes. Then the first equation 
 \bear
-\frac{3}{2} \zeta  F_0(\zeta ) {dF_0(\zeta )\over d\zeta}-{3\over2}\zeta F_0(\zeta)+3 (F_0(\zeta ))^2+3 F_0(\zeta )+\frac{2}{3}=0
\label{eq:F0a}
\eear
 determines $F_0(\zeta)$ to be 
 \bear
F_0(\zeta)&=&  \frac{1}{6} \left(-2+c^2 \zeta ^2\pm  c\, \zeta\, \sqrt{4+ c^2\zeta^2}\right)\approx -{1\over3}\pm {c\over 3}\zeta+\dots
\label{eq:F0b}
\eear
 where $c$ is an integration constant. Matching with the trans-series expansion (\ref{eq:trans-ansatz}) fixes
  $c=3$ and selects the upper sign:
 \bear
 F_0(\zeta)=-{1\over3}+{3\over 2}\zeta^2 + \zeta \sqrt{ 1+{9\over 4} \zeta^2}
 \label{eq:F0c}
 \eear
 From the small $\zeta$ expansion of $F_0(\zeta)$ one can read off the leading (i.e. $w^0$) coefficient of the $n^{th}$ non-hydrodynamic series. To see this explicitly, compare
 \bear
 F_0(\zeta)&=&-{1\over3}+\zeta+{3\over2}\zeta^2+{9\over8}\zeta^3+\dots\nonumber\\
 &\equiv & f^{(0)}_0+f^{(1)}_0\,\zeta+f^{(2)}_0\,\zeta^2+f^{(3)}_0\,\zeta^3+\dots
 \label{eq:F0}
 \eear
 with the leading coefficients given in Eqs. \eqref{eq:late1}, \eqref{eq:f1},  \eqref{eq:f2} and  \eqref{eq:f3}, with $C_\lambda=2C_\eta$.
 
 The next equation in the tower (\ref{eq:tower}) determines $F_1(\zeta)$
 \bear
-{3\over2}\zeta\left( F_0+1\right) {dF_1\over d\zeta}+3\left(2F_0+1\right)F_1
-3C_\eta\,\zeta\left(F_0+1\right){dF_0\over d\zeta}
+4C_\tau F^2_0+{8\over3}C_\tau F^2_0
+\frac{4}{9} (C_\tau-C_\eta)=0\nn
\label{eq:F1a}
\eear
Note that this equation is {\it linear} in $F_1(\zeta)$.
After choosing the integration constant to match the $1/w$ term of the first non-hydrodynamical expansion \eqref{eq:f1}, we obtain
\bear
F_1(\zeta)={\left(3\zeta+\sqrt{4+9\zeta^2}\right)^2\over18 C_\tau \sqrt{4+9\zeta^2}}&&\left[54 C_\eta(C_\tau-C_\eta) \zeta+ C_\tau\left(9(C\tau-C\eta)\zeta^2+2C_\eta\right)\left(-3\zeta+\sqrt{4+9\zeta^2}\right)\right. 
\nn
&& \left. +6C\tau(2C_\tau-3C\eta)\,\zeta \sinh^{-1}\left({3\zeta\over2}\right)\right]
\label{eq:F1b}
\eear 
Similarly to \eqref{eq:F0}, the small $\zeta$ expansion of $F_1(\zeta)$ generates all the $f^{(n)}_1$ coefficients in the trans-series:
\bear
F_1(\zeta)&=&{4C_\eta\over9}+{2\,C_\eta \over 3}\left( 10-{9\,C_\eta\over C_\tau}\right)\zeta+\left(13 C_\eta-{18 C_\eta^2\over C_\tau}+4C_\tau \right)\zeta^2\nn
&&+\left({15 C_\eta\over 2}-{81 C_\eta^2\over 4\,C_\tau}+9C_\tau \right)\zeta^3+\dots\nn
&=& f^{(0)}_1+f^{(1)}_1\,\zeta+f^{(2)}_1\,\zeta^2+f^{(3)}_1\,\zeta^3+\dots
\label{eq:F1c}
\eear
which can be checked from Eqs. \eqref{eq:late1}, \eqref{eq:f1},  \eqref{eq:f2} and  \eqref{eq:f3}, with $C_\lambda=2C_\eta$. This procedure can be continued to determine all the $F_k(\zeta)$, as all the equations are linear for $k\geq 1$.
\begin{figure}[h]
\includegraphics[scale=0.4]{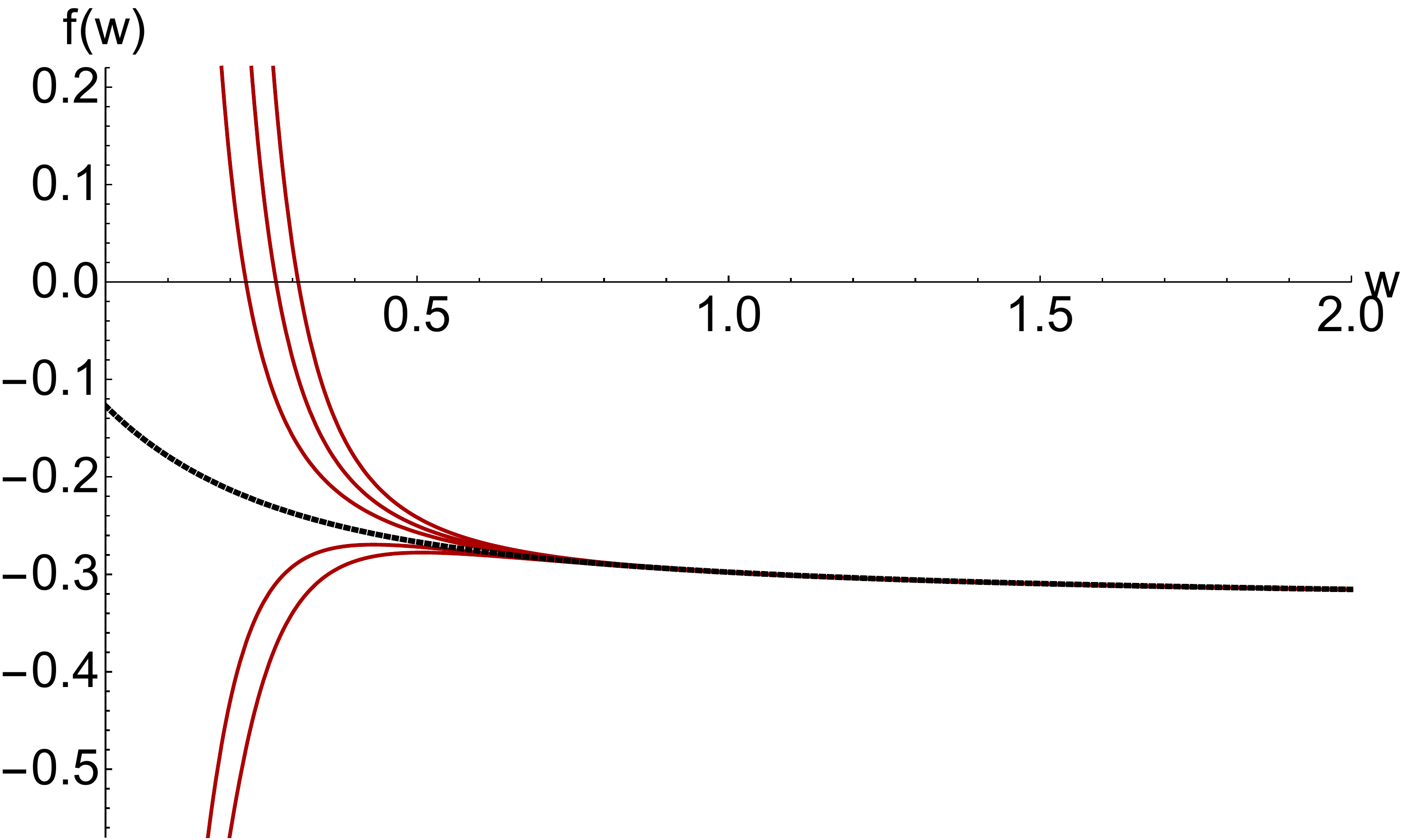}
\caption{The black (dashed) line is the numerical integration of the nonlinear differential equation (\ref{eq:main-eqn}), starting at very early time $w=10^{-6}$. The red (solid) lines correspond to the trans-asymptotic expansion (\ref{eq:transasymptotic}), using the expressions for $F_0$ and $F_1$ in (\ref{eq:F0c}, \ref{eq:F1b}), the imaginary part of the trans-series parameter $\sigma$ fixed by (\ref{eq:stokes}, \ref{eq:canc}), and different values of the real part of $\sigma$ (from bottom to top: $-0.2,-0.12,0.06,0.12,0.2$). This clearly illustrates that the physical meaning of the real part of the trans-series parameter $\sigma$ is that it corresponds to matching a particular member of the family of trans-series to a particular solution with a prescribed initial conditon. A detailed example of such a matching procedure for the Painlev\'e I equation can be found in \cite{costin-dubrovin}.}
\label{fig:trans}
\end{figure}

The advantage of this procedure can be seen when  we reintroduce the trans-series parameter $\sigma$ factor in the argument of $F_k$, writing again $F_k(\sigma\, \zeta)$. The condition of reality of $f(w)$ fixes the imaginary part of $\sigma$ in terms of the (numerically determined) Stokes parameter $S_1$ in equation (\ref{eq:stokes}, \ref{eq:canc}), 
but the real part of $\sigma$ is arbitrary \cite{costin-costin,Aniceto:2013fka}. This remaining real constant parametrizes the entire family of trans-series solutions to the differential equation, each member  of which connects to the hydrodynamical expansion at large $w$, but which have different behavior once $w$ becomes small enough that the exponential factors become comparable with powers of $w$. Thus, this real part of the trans-series parameter characterizes the various different possible ``initial conditions'', at early times (i.e. small $w$), each of which eventually tends to the hydrodynamical expansion as $w\to\infty$. In \cite{Heller:2015dha} this phenomenon was referred to as an ``attractor''. This is in fact a generic feature of the large class of nonlinear differential equations whose solutions can be expressed in trans-series form \cite{costin,costin-costin}. This behavior is illustrated in Figure \ref{fig:trans}, in which we vary the real part of $\sigma$, and see that this corresponds to different possible choices of initial condition. This simple observation resolves the apparent inconsistency between the fact that the formal hydrodynamic series expansion (\ref{eq:late1}) has no undetermined parameter, while the exact (numerical) solution to the first-order nonlinear equation (\ref{eq:main-eqn}) clearly depends on one initial condition: this is why there is a family of trans-series solutions, parametrized by the trans-series parameter $\sigma$, more precisely by its real part if $f(w)$ is real.

\section{Conclusions}

We have shown that the nonlinear ordinary differential equation describing boost-invariant  Bjorken flow in conformal hydrodynamics exhibits characteristic signs of resurgence. The formal late-time Taylor expansion of the hydrodynamic derivative expansion is asymptotic, and can be systematically extended to a trans-series expansion, for which we have characterized the Borel plane structure. This analysis reveals  precise relations between the fluctuations about different  ``non-perturbative sectors'', associated with modes beyond hydrodynamics. These modes are exponentially damped and the damping rate, or the inverse relaxation time, plays the role of an  instanton action in the language of semi-classical physics. The factorial large order  growth of the fluctuation coefficients in one sector is precisely determined by low orders of expansions about neighboring sectors. This means, in particular, that the 
asymptotic hydrodynamic expansion encodes physical information about non-hydrodynamic modes.
For example, the trans-series expansion of the energy density $\E(\tau)$ at late proper-time includes also exponentially small terms $\sim \tau^\alpha \exp\left[- c\, \tau^{2/3}\right]$, and these non-hydrodynamic modes have an interpretation in terms of quasi-normal modes in the gravitational picture. Resurgent asymptotic analysis suggests that these general features should extend to the partial differential equations in more general hydrodynamical problems with gravity duals, in which the metric coefficient functions are expanded in terms of both proper time and distance from the horizon.

\vskip0.2cm
{\bf \large Acknowledgements }
\vskip0.2cm
We gratefully acknowledge discussions with  Ovidiu Costin, Michal Heller, Michal Spalinski, Dam Son, Misha Stephanov, and Ho-Ung Yee. This work was supported by the U.S. Department of Energy under Contracts No. DE-FG02-93ER-40762 (GB) and DE-SC0010339 (GD).


\end{document}